\documentclass[letterpaper,12pt]{article}

\usepackage{ulem}
\usepackage[authoryear,comma,longnamesfirst,sectionbib]{natbib}

\usepackage{amssymb}
\usepackage{amsmath}
\usepackage{graphicx}
\usepackage[final]{pdfpages}
\usepackage[margin=0.0cm]{caption}
\usepackage{picins}

\usepackage{enumitem}\setlist[description]{leftmargin=\parindent,labelindent=\parindent}

\usepackage{hyperref}
\usepackage{keyval}

\usepackage{fullpage}

\definecolor{lightblue}{rgb}{.80,1,1}
\definecolor{myblue}{rgb}{.5,.5,.8}
\definecolor{dblue}{rgb}{0,.5,1}
\definecolor{lightpurple}{rgb}{1,.80,1}
\definecolor{lightyellow}{rgb}{1,1,.70}
\definecolor{webgreen}{rgb}{0,.5,0}
\definecolor{lightgrey}{rgb}{.8,.8,.8}

\newtheorem{definition}{Definition}

\newcommand{\real}{{\mathbb R}}

\newcommand{\reald}{\real^{\rm{d}}}

\newcommand{\n}{{\mathbb N}}

\begin{document}

\title{\bf Using persistent homology and dynamical distances to analyze protein binding}

\author{Violeta Kovacev-Nikolic
\thanks{Department of Mathematical and Statistical Sciences, University of Alberta, Canada.
Email: violeta.kovacev-nikolic@ualberta.ca}
,\
Peter Bubenik
\thanks{Department of Mathematics, Cleveland State University, USA. Email: peter.bubenik@gmail.com}
,\
Dragan Nikoli\'c
\thanks{Department of Mechanical Engineering, University of Alberta and National Institute for Nanotechnology, Canada.
Email: dnikolic@ualberta.ca}
,\
Giseon Heo
\thanks{Corresponding author. School of Dentistry; Department of Mathematical and Statistical Sciences, University of Alberta, Edmonton, Canada, T6G 2N8.
Email:  gheo@ualberta.ca}
}

\date{July 23, 2015}

\maketitle

%
%

\begin{abstract}
\noindent
Persistent homology captures the evolution of topological features of a model as a parameter changes.
{
The most commonly used summary statistics of persistent homology are the barcode and the persistence diagram.
Another summary statistic, the persistence landscape, was recently introduced by Bubenik.
}
It is a functional summary, so it is easy to calculate sample means and variances, and it is straightforward to construct various test statistics.
Implementing a permutation test we detect conformational changes between closed and open forms of the maltose-binding protein, a large biomolecule consisting of 370 amino acid residues.
 Furthermore, persistence landscapes can be applied to machine learning methods. A hyperplane from a support vector machine shows the clear separation between the closed and open proteins conformations.
Moreover, because our approach captures dynamical properties of the protein our results may help in identifying residues susceptible to ligand binding;
we show that the majority of active site residues and allosteric pathway residues are located in the vicinity of the most persistent loop in the corresponding filtered Vietoris-Rips complex.
This finding was not observed in the classical anisotropic network model.
\end{abstract}

%
%
\vspace{0.4in}


%
%
\newpage
\section{INTRODUCTION}
\label{Sec:Intro}

In this paper, we describe topological techniques for the analysis of geometric data.
In particular we apply these methods to study a particular protein, the maltose-binding protein (MBP), whose geometric shape
can be represented by 370 points in $\mathbb{R}^3$, or equivalently, as one point in $\mathbb{R}^{3 \times 370}$.
However, this structure is not static; it is dynamic.
It ``jiggles'' under thermal fluctuations, and changes among various conformations as it performs its biological functions.
The space of all such conformations is a subspace of $\mathbb{R}^{3\times 370}$.
We use topology to construct summary statistics of these conformations and see what they can tell us about our data.
Instead of working with the static spatial coordinates, we use a richer dynamic model of the protein from which we use correlations to
calculate a $370\times 370$ matrix of dynamical distances for each conformation.
This matrix is the input for our topological and statistical methods.

{
Since our computational protocol only requires the knowledge of the dynamic cross correlation, it can also be used in reducing the massive
amount of data generated by long-term molecular dynamics (MD) simulations of large molecular systems of arbitrary complexity.
For example, with help of effective pair potentials obtained from the DRISM-KH molecular theory of solvation \citep{Kobryn:2014} further speed-up
of MD simulations is possible at various levels of coarse-grained mapping.
For given force-field parametrization and under physiological conditions, the time ensemble average of atomistic details of communications among separate
regions of macromolecules often exhibit multi-modal behavior.
Modal decomposition of dynamic cross correlation and its convergence analysis for the present study is detailed in \cite{MBP:DynamicalModel}.
Briefly, in constructing the matrices of dynamical distances, the reference structures we use are the experimental crystal structures reported in Table~\ref{Tab:MBPstructures}.
In conventional uni-modal MD simulation the reference structure can be iteratively updated from averaged coordinate of trajectory segments, however the explicit inclusion of multi-modal motions of atoms is also possible \citep{Kasahara:2014}.
Spatially perturbed reference structures detected along the simulated trajectory are expected to yield the biologically important conformational changes only if they alter the dynamic cross correlation matrix in such way to result in a significant rearrangement of the residues that define the topologically most persistent loop.
}

Algebraic topology classifies topological spaces using topological invariants such as homotopy, homology, and Euler characteristic.
Homology detects topological features such as connected components, holes and voids.
Computational topologists have introduced a variation of homology called \emph{persistent homology} which records the history of appearances and disappearances of topological features as a parameter changes (\citet{Edels2002}, \citet{ZomCar2005}).
Here is a basic example of persistence. Given a set of points sampled from some (unknown) object in a Euclidean space, consider a set of balls centered at these points with a common radius.
For certain radii, this union of balls will approximate the unknown object in various ways.
Persistent homology summarizes the appearance and disappearance of topological features of the union of balls as the radius increases.
Topological attributes that persist over a large range of radii are likely to be a signal while short-lived ones are likely to be noise.
Three usual summaries of persistent homology are the \emph{barcode} \citep{CZCG2004a}, the \emph{persistence diagram} \citet{Edels2002}, and the \emph{persistence landscape} \citep{Bubenik2015}.
The space of barcodes or persistence diagrams is a space in which it is difficult to do statistics, though theoretical advances have been made by \citet{mileyko2011} and \citet{Turner2013}, as well as \citet{HGK2012} who applied statistics to landmarks of maxillary area in an orthodontic study.
However, the space of persistence landscapes has the advantage of being a vector space (in fact, a separable Banach space).
\citet{ChazalEtAl2013} and \citet{ChazalEtAl2014} have studied the weak convergence, convergence of the bootstrap, and confidence bands for the average persistence landscape.
Furthermore, as observed by \citet{Reininghaus2014}, the persistence landscape may be viewed as the feature map of a positive definite kernel, allowing one to apply all of the usual kernel methods of machine learning.
\citet{Reininghaus2014} also developed an alternative kernel-based persistence approach.
In this article we will use the persistence landscape for our statistical data analysis,
{
a more detailed discussion of which can be found in an early study \citep{MScThesis} of twelve samples of the HIV-1 protease.
}
We also show how the barcode may be useful in exploratory data analysis.
For homology group of degree 1, the longest interval in the barcode corresponds to the most persistent loop in the data.
{ In the present study} we show that biologically pertinent units, the active sites and the allosteric pathway residues, are located close to this loop.
{ This is an important information that may guide the subsequent docking studies of protein affinities to numerous drug fragments under experimentally relevant conditions as previously done for prion protein and thiamine \citep{Nikolic:2012}.}

{ For recent applications of persistent homology to studies of protein structure, folding, and flexibility we refer the reader to the articles \citep{Xia:2014, Xia:2015a, Xia:2015b}.
}
{[A14]
Apart from the use of \textsc{javaPlex} library \citep{javaPlex} to visualize the associated barcodes, there is no further similarity between the present or the past \citep{MScThesis} topological analyses and the recent method of molecular topological fingerprints by \cite{Xia:2014}.
They used Betti numbers to study protein structure, folding, and flexibility.
An interesting aspect of \cite{Xia:2015a} article is that Betti numbers are calculated at various scale of other parameters.
The other parameters are configuration index, cut-off distance, scale of resolution, and number of iterations.
}

A brief outline of the article is given as follows.
In Section~\ref{Sec:MBP} we explain our biological motivation and introduce basic facts concerning the MBP and describe the data.
Section~\ref{Sec:ComputationalProtocolandData} lists the main steps of our topological data analysis.
We define the persistence landscape \citep{Bubenik2015} from which we derive a suitable random variable; this allows us to formulate statistical hypotheses of interest and construct a test statistic.
In Section~\ref{Sec:ResultsAndDiscussion} we compare open and closed conformations of MBP and extract additional information from visualized complexes.
Our conclusion and signposts for future work appear in Section~\ref{Sec:Summary}.

%
%
\section{THE PROTEIN DATA}
\label{Sec:MBP}


The maltose-binding protein is a bacterial protein found in \emph{Escherichia coli} where its primary function is to bind and transport sugar molecules across cell membranes, providing energy to the bacterium \citep{BoosShuman:1998}.
Though sometimes causing serious illness, most strains of \emph{E. coli} are nonpathogenic and in fact beneficial.
These bacteria colonize the gastrointestinal tract of humans and animals and protect the gut from harmful bacteria \citep{Hudault:2001}.
Furthermore, \emph{E. coli} is the best known living organism and is used to study various cellular processes \citep{VanHoudt:2005}.
In our paper we use topological methods to study the MBP.

While performing biological functions, the MBP changes its structure.
Our objects of study are fourteen three-dimensional structures of the MBP, obtained by X-ray crystallography.
The structures are available from the Protein Data Bank, \citet{PDB}.
Each structure is a large biomolecule with about 3000 heavy atoms grouped into 370 relatively small clusters representing amino acid residues.
The term \emph{residue} refers to the fact that during the formation of the protein the constituent amino acids lose parts of their water molecules \citep{IUPAC}.  Instead of using an all-atom model, we use a computationally more affordable \emph{coarse-grained model} in which every residue is represented by a single unit
\citep{Cavasotto:2005}.
{
This is compatible with the topological approach as it reduces local information but preserves global topological information.
}


%
\begin{figure}[!htbp]
  \begin{center}
  \resizebox{\columnwidth}{!}{

\includegraphics[angle=0.,width=0.475\textwidth]{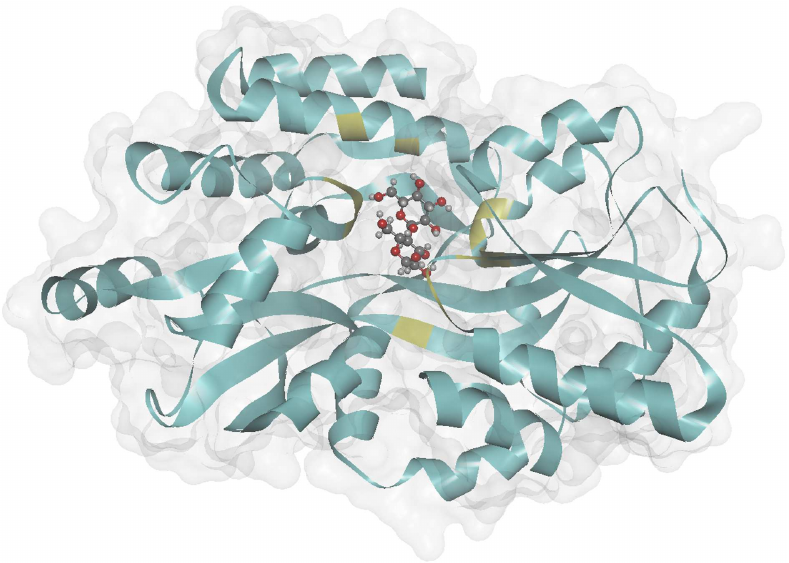}
\label{Fig:1mpd_bio_r_500}
       \includegraphics[angle=0.,width=0.475\textwidth]{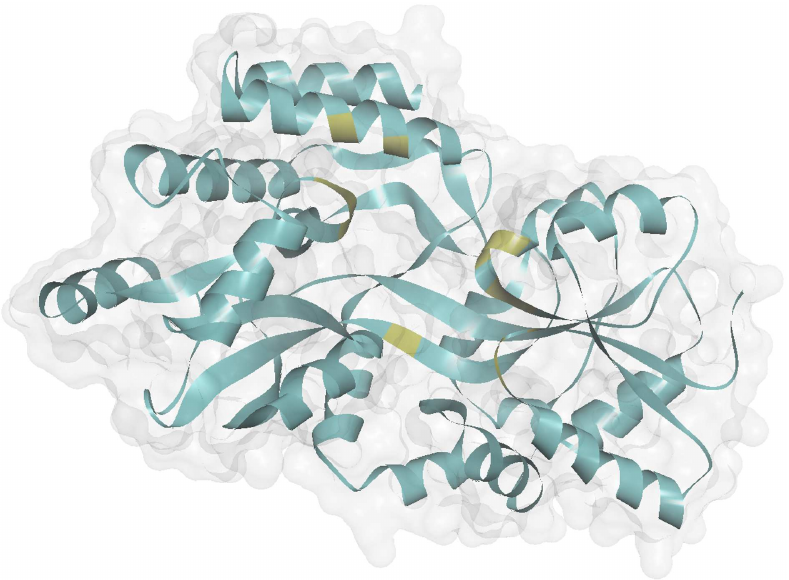}
\label{Fig:1omp_bio_r_500}
          }
  \end{center}
  \caption{The biological assembly for the closed-holo 1MPD conformal structure (left, \citet{Shilton:1996}) and the open-apo 1OMP conformal structure (right, \citet{Sharff:1992}). {  Secondary structures and solvent accessible surfaces of both proteins are shown as blue flat ribbons and gray transparent surfaces, respectively. Active sites in ribbon representations have yellow color and interact with ligand maltose shown here as ball and stick model embedded in 1MPD structure.}}
  \label{Fig:MBP_BiologicalAssembly}
\end{figure}

A major conformational change in the protein occurs when a smaller molecule called a \emph{ligand} attaches to the protein molecule, see Figure~\ref{Fig:MBP_BiologicalAssembly}.
\citet{Szmelcman:1976} determined that the MBP interacts with various sugar molecules (ligands), starting from the small maltose molecule through the larger maltodextrin.
Ligand-induced conformational changes are important because the biological function of the protein occurs through a transition from a ligand-free (\emph{apo}) to a ligand-bound (\emph{holo}) structure \citep{Seeliger:2010}.

Simulations and to some extent experiments show that 95\% of the time the two domains of MBP are separated and twisted, which is called an \emph{open} conformation, and 5\% of the time they are close to each other, which is called a \emph{closed} conformation.
If closed, it is always due to having a captured ligand, see Figure~\ref{Fig:MBP_BiologicalAssembly}.
Open structures can have an attached ligand or not, as verified by experiments (see Table~\ref{Tab:MBPstructures}).
From a practical viewpoint, closed conformations are more significant because they are more stable so the detachment of the bound ligand is less likely to happen.

We consider seven closed and seven open conformations.
We differentiate between them using deformation energies; the energetically more favorable closed conformation binds a sugar molecule while its open counterpart requires greater deformation energies and may or may not engage in the binding process.
The list of the fourteen MBP structures we investigate is shown in Table~\ref{Tab:MBPstructures}.
\begin{table}[!htbp]\small
\centering
\caption{Fourteen MBP structures with the names of the ligands for \emph{holo}-forms. Each structure is labeled by a four-letter Protein Data Bank \citep{PDB} code. \label{Tab:MBPstructures}}
    \begin{center}
    \resizebox{0.75\columnwidth}{!}{ 
        \begin{tabular}{ccllcc}
            No.& PDB code & Ligand name && Protein structure & Reference \\
            \hline
            1 & 1ANF & maltose       && closed-{\em holo} & \citet{Quiocho:1997} \\
            2 & 1FQC & maltotriotol  && closed-{\em holo} & \citet{Duan:2001} \\
            3 & 1FQD & maltotetraitol&& closed-{\em holo} & \citet{Duan:2001} \\
            4 & 1MPD & maltose       && closed-{\em holo} & \citet{Shilton:1996} \\
            5 & 3HPI & sucrose       && closed-{\em holo} & \citet{Gould:2010} \\
            6 & 3MBP & maltotriose   && closed-{\em holo} & \citet{Quiocho:1997} \\
            7 & 4MBP & maltotetraose && closed-{\em holo} & \citet{Quiocho:1997} \\ \hline
            8 & 1EZ9 & maltotetraitol&& open-{\em holo}   & \citet{Duan:2002} \\
            9 & 1FQA & maltotetraitol&& open-{\em holo}   & \citet{Duan:2001} \\
            10& 1FQB & maltotetraitol&& open-{\em holo}   & \citet{Duan:2001} \\
            11& 1JW4 &  -            && open-{\em apo}    & \citet{Duan:2002} \\
            12& 1JW5 & maltose       && open-{\em holo}   & \citet{Duan:2002} \\
            13& 1LLS &  -            && open-{\em apo}    & \citet{Rubin:2002} \\
            14& 1OMP &  -            && open-{\em apo}    & \citet{Sharff:1992} \\ \hline
          \end{tabular}
        }
    \end{center}
\end{table}

Using professional docking software like Discovery Studio by Accelrys, one can only visually determine whether or not a conformal structure can be classified as
open or closed.
An alternative approach, based on modeling of deformation energies in protein structures \citep{MBP:DynamicalModel}, can also differentiate between the two groups in a more systematic way.
In this paper we will show yet another approach that relies on topological/statistical methods.
Note that out of 370 residues, fewer than twenty are actively involved in sugar binding.  These residues which are crucial for the function of the
protein are referred to as \emph{active sites}. However, identifying active sites inside a protein structure is difficult.
Active sites can be identified using experimental methods that engineer different parts of the protein to be preferential in binding
ligands; or theoretical methods can be used to model the various physical interactions between atoms of the protein and the ligand \citep{Amitai:2004}.


We also consider \emph{allosteric pathway residues} which behave as bridges between the ligand binding site and the exterior of the protein \citep{Lockless:1999}. Due to thermal vibrations, a closed-holo structure may easily transform to an open-holo in which the risk for detachment of the bound sugar molecule is higher. The stability of the closed form can be increased by influencing the active sites.
In a closed conformation they are deeply buried in the interior of the protein and inaccessible to direct influences, however, an indirect access is feasible through the effect of \emph{allostery} \citep{Rizk:2011}.


\subsection{Dynamical distances}
\label{Subsec:OurData}

One {may} attempt to use topological methods to describe the function of the MBP using the spatial coordinates of the residues.
This is not a novel idea; \citet{EHbook2010} studied computational ways of predicting protein interactions based solely on their shape.
Furthermore, \citet{Gameiro:2012} defined a topological quantity based on persistence diagrams of several proteins and established correspondence with experimental compressibility of the majority of investigated proteins.
However, this intuitive approach proves to be inefficient in distinguishing between the closed and the open MBP conformation.

The three-dimensional coordinates obtained by x-ray crystallography give a snapshot of the conformal structure.
However, this structure is really time dependent and wobbly.
{
The analysis by x-ray diffraction requires samples to be in solid state, which is quite different physico-chemical phase than the physiological solution found \emph{in vivo}.
Namely, to facilitate and optimize protein crystallization various organic solvents, polymers, and salts are typically used as precipitants -- none of which are present \emph{in vivo}.
Moreover, the atom positions in x-ray crystallography are deduced from localized electron density maps while in complex physiological solutions inter-atomic distances in proteins are
influenced by constraints imposed by their interaction with highly dynamical solvent molecules.
}
So a dynamic descriptor is more appropriate.
Therefore, we do not analyze the geometry of the MBP structure directly.
Instead we use the static crystallographic data to construct a dynamic model of the protein structure from which we calculate \emph{dynamical distances} between the residues.
Our subsequent analysis will use these dynamical distances and not the geometric distances.


We model the dynamics of the protein structure using an \emph{elastic network model} \citep{Atilgan:2001,Tobi:2005}.
Though all constituents of the protein constantly exhibit small oscillations due to thermal motion, movements on larger scale occur because neighboring units strongly affect each other.
Hence the motion and function of the biomolecule are the result of the coordinated action of mutually interacting residues, i.e. the protein is modeled like a
dynamical system of beads joined by elastic springs
{
with the cut-off distance of 15{\AA} (see Fig.~SM-4 of \cite{MBP:DynamicalModel}). We note that main results of our topological/statistical analysis for all investigated structures remain
robust against cut-off distances larger than 4{\AA}, become well-converged at about 12{\AA}, and numerically insensitive above 20{\AA}.
}
An energy state of such a molecule is a superposition of normal modes of oscillations, leading to different spatial conformations depending on the deformation energy.
There exist infinitely many energy modes; however, as a compromise between numerical accuracy and computational efficiency, our model takes into account the lowest twenty nontrivial energy modes \citep{MBP:DynamicalModel}.
{
This value is obtained as the lowest normal mode for which the averaged difference in deformation energies between open and closed structures remains constant (see Fig.~SM-5 of \cite{MBP:DynamicalModel}).
}

\begin{figure}[!htbp]
   \begin{center}
   \resizebox{\columnwidth}{!}{
       \includegraphics[angle=0.,width=1.0\textwidth]{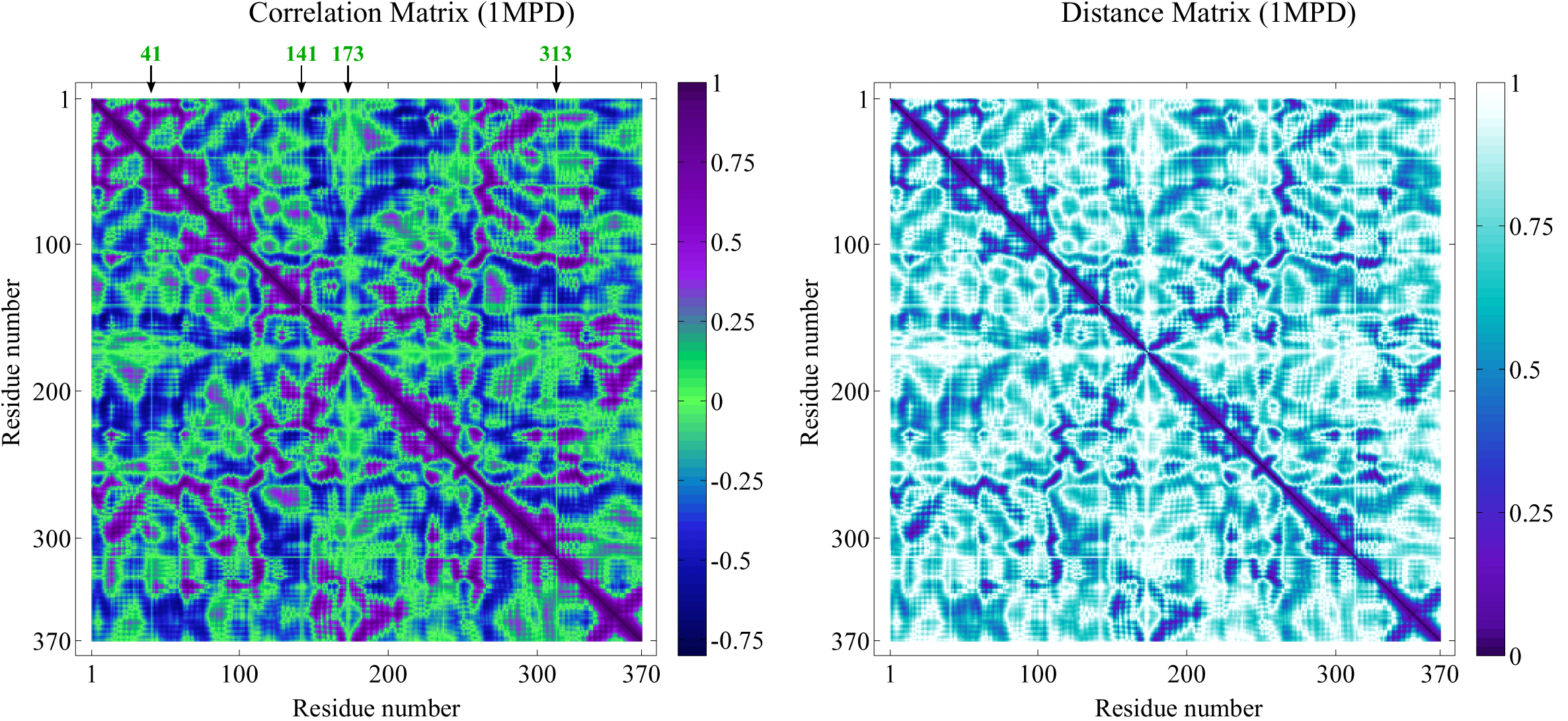}
           }
   \end{center}
       \caption{Visualization of the cross-correlation matrix and the dynamical distance matrix of the 1MPD structure. Axes correspond to residue indices running from 1 to 370.
       (Left) Cross-correlation matrix for the 1MPD structure. Dark regions correspond to high pairwise correlations, with -0.76 as the most negative value. Green horizontal and vertical lines correspond to most flexible residues which poorly correlate with the rest of the protein structure thus their total correlations are approximately zero.
       (Right) Dynamical distance matrix for the 1MPD structure,
       calculated from the correlation matrix using
       Equation~\eqref{Eq:DistanceMatrix}; the linear relationship
       causes a similar visual layout of the two matrices.
     }
\label{Fig:1MPD_CorrelationAndDistanceMatrix}
\end{figure}

Taking into account the first twenty nontrivial modes of oscillation, we calculate fourteen cross-correlation matrices ($\mathbf{C}$) of size $370 \times 370$ using the Anisotropic Network Model web server \citep{ANM:2006}.
Following \citet{Bradley:2008}, for each correlation matrix we calculate the associated \emph{dynamical distance} matrix ($\mathbf{D}$) using a simple linear map, \begin{equation}\label{Eq:DistanceMatrix}
    D_{ij} = 1 - \left| C_{ij} \right|,
\end{equation}
though other choices are also possible.
This defines a metric space in which highly (anti)correlated residues lie close to each other.
An illustration of the matrix $D$ for the $1MPD$ structure is given in Figure~\ref{Fig:1MPD_CorrelationAndDistanceMatrix}.


\begin{figure}[!htbp]
   \begin{center}
   \resizebox{0.95\columnwidth}{!}{
      \includegraphics[angle=0.,width=0.95\textwidth]{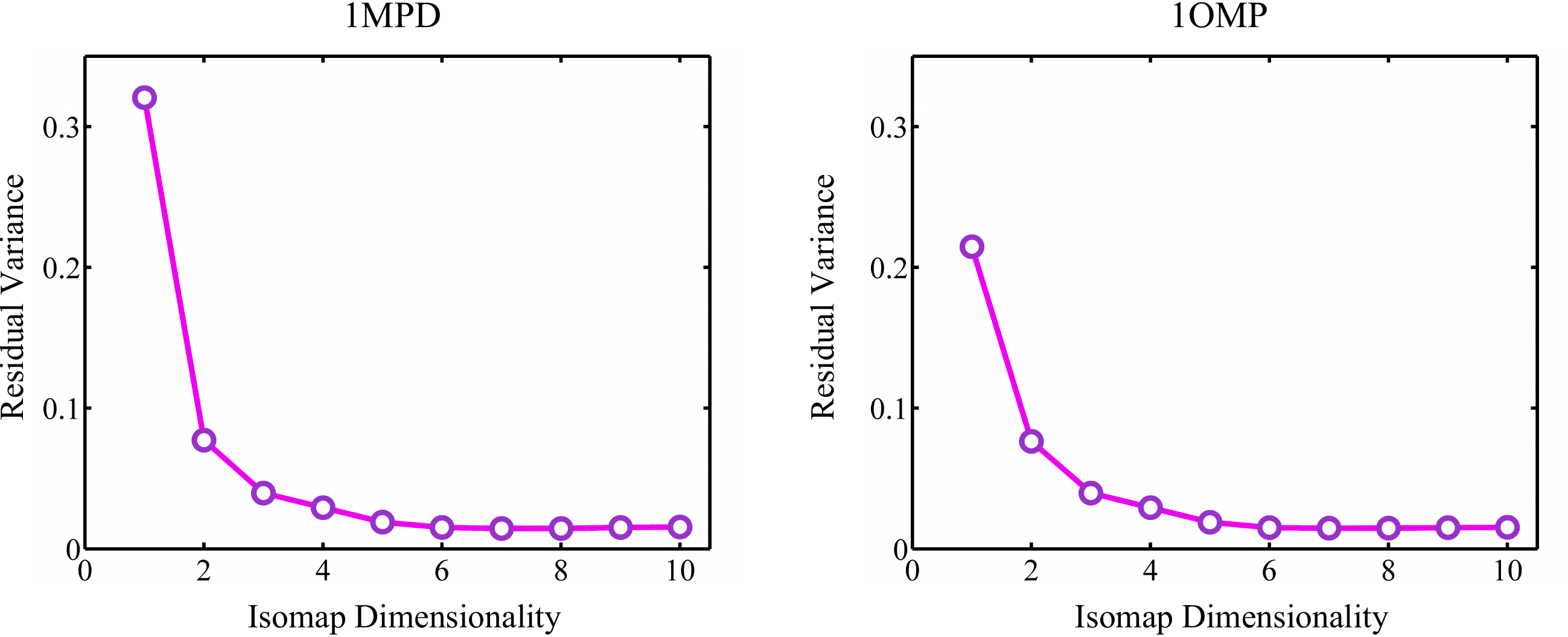}
}
   \end{center}
   \caption{Dimension reduction via \textsc{Isomap} for the 1MPD (left) and the 1OMP (right) structure: the `elbows' in scree plots suggest that a three dimensional embedding is appropriate. }
   \label{Fig:MBP_ScreePlots}
\end{figure}

To visualize this metric space we apply the nonlinear dimension reduction method, ISOMAP \citep{TenenbaumISOMAP}.
We prefer this method to {\bf MDS} because for our data the errors on distances are smaller.
As indicated by the scree-plots in Figure~\ref{Fig:MBP_ScreePlots}, projecting to three dimensions is appropriate.


%

%
%
\section {TOPOLOGICAL METHODS}
\label{Sec:ComputationalProtocolandData}

In this section we outline how topological methods can be applied to geometric data and how these tools can be combined with statistical analysis.
A $n \times n$ distance matrix $\mathbf{D}$ defines a discrete metric space on $n$ points $x_1,\ldots,x_n$, where $d(x_i,x_j) = D_{ij}$.
From this we construct a parametrized family of simplicial complexes.
Given $d \geq 0$, let $\mathcal{R}_d$ denote the simplicial complex on $n$ vertices $x_1,\ldots,x_n$, where an edge between the vertices $x_i$ and $x_j$ with $i\neq j$ is included if and only if $d(x_i,x_j) \leq d$; more generally, we include the $k$-simplex with vertices $x_{i_0},\ldots,x_{i_k}$ if and only if all
of the pairwise distances are at most $d$.
This simplicial complex is called a \emph{Vietoris-Rips complex}.
Since for $d \leq d'$, $\mathcal{R}_d \subseteq \mathcal{R}_{d'}$ this family is a filtered simplicial complex.
Notice that there are only finitely many values of $d$ for which we obtain a distinct simplicial complex. For computations we restrict to this finite filtration.

Of interest is the topology of this simplicial complex and how it changes as the parameter changes.
In particular we are interested in $H_k(\mathcal{R}_d)$ the homology of the Vietoris-Rips complex with coefficients in the field $\mathbb{Z}/2\mathbb{Z}$, for small values of $k$.
For coefficients in a field, homology is a vector space.
For $k=0$ this vector space has as a basis the connected components of the simplicial complex.
For $k=1$ its basis consists of linearly independent cycles that are not boundaries.
Similar statements hold in higher degrees.
More details can be found in \citet{Hatcher}, for example.
For $d \leq d'$, the inclusion $\mathcal{R}_d \subseteq \mathcal{R}_{d'}$ induces a linear map $H_k(\mathcal{R}_d) \to H_k(\mathcal{R}_{d'})$.
The set of vector spaces $\{\mathcal{R}_d\}$ together with the corresponding linear maps is referred to as a \emph{persistence module}.
This persistence module can be completely described by a collection of intervals referred to as a \emph{barcode} (\citet{Edels2002,ZomCar2005}).
By representing each interval by its endpoints, one obtains a collection of points in $\mathbb{R}^2$, called a \emph{persistence diagram}.
{
 The persistence module cannot be recovered from the Betti numbers $\{\dim H_k(\mathcal{R}_d) \}_d$ since this provides no information on the linear maps.
}
Various software packages compute barcodes; we use the \textsc{javaPlex} library \citep{javaPlex}.
The distance matrix may be obtained from the Euclidean distances between a collection of points in $\reald$ (point cloud data), the diffusion metric (\citet{Bendich2011}), similarity, correlation and covariance matrices.
Filtered simplicial complexes can also be obtained from Morse functions and kernel estimators (\citet{Bubenik2010}).


\begin{figure}[!hbtp]\small
 \begin{center}
 \resizebox{0.95\columnwidth}{!}{
\includegraphics[angle=0,width=0.60\textwidth]{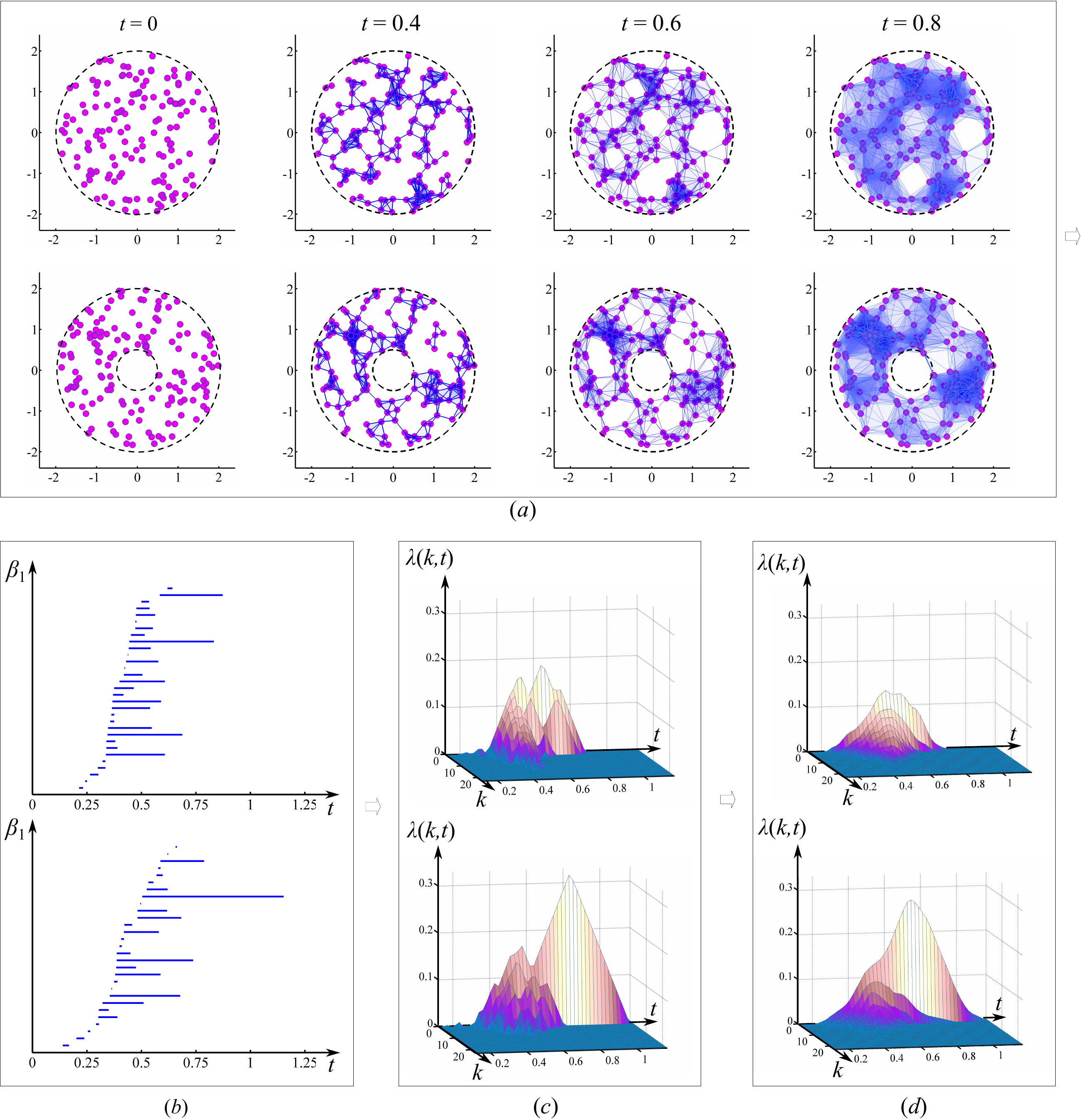}
   }
 \end{center}
 \caption{Steps of our topological statistical analysis of points sampled for a disk and an annulus. (a) Three snapshots of the two
   filtered Vietoris-Rips complexes. Initially we sample point cloud data from a disk and an annulus; as the filtration parameter
   increases points get connected, different loops are born and die. (b) The birth and death time of loops is recorded in a
   barcode for the first homology  $H_1$. The long bar in the case of the annulus detects the hole, whereas the shorter (`noisy') bars
   for the disk and annulus detect transient phenomena. (c) The persistence landscape (PL) corresponding to each barcode.
(d) The mean persistence landscapes for the 10 disks and the 10  annuli. A
   permutation $t$-test ($p$-value $0.0028$) differentiates the disk from the annulus in terms of the one-dimensional cycle, that is,
   the loop. }
   \label{Fig:FlowDiskAnnulus}
\end{figure}

Consider the following simple example, illustrated in Figure~\ref{Fig:FlowDiskAnnulus}.
We sample 150 points uniformly and randomly from a circle and annulus.
From such point cloud data (PCD) we construct the corresponding filtered Vietoris-Rips complex and calculate the associated barcode.
When $d = 0$ we have just the points; when $d = 0.4$ several small loops appear; later when $d = 0.8$ most loops disappear and only a few remain.
The time of appearance and disappearance of loops is recorded in a barcode for the first homology group $H_1$, see Figure~\ref{Fig:FlowDiskAnnulus}(b).
A barcode consists of intervals that indicate the times of birth (starting points of lines), death (end points of lines), and the duration of survival (lengths of lines), of a topological feature (a loop, in this example).
In the next section we will apply this construction to fourteen conformations of the maltose-binding protein.
Recall that in our model every residue is represented by a single point; using the spatial structure we would get 370 points in $\mathbb{R}^3$.
Using dynamical distance we have a $370 \times 370$ distance matrix.
In either case we can calculate the corresponding barcodes.

\begin{figure}[!htbp]
 \begin{center}
 \resizebox{\columnwidth}{!}{
  \includegraphics[angle=0,width=0.9\textwidth]{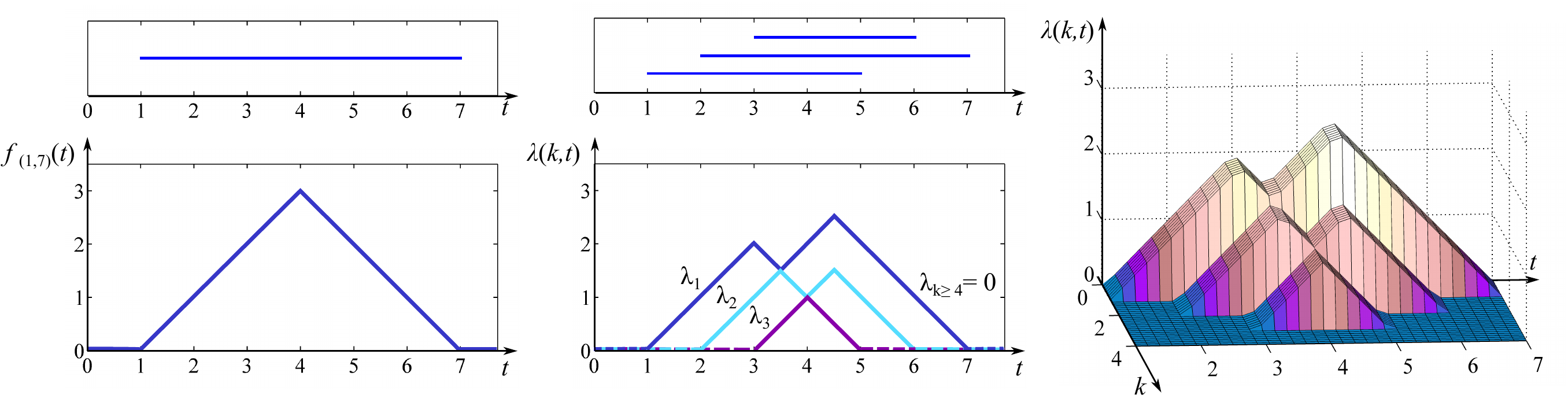}
   }
 \end{center}
 \caption{Construction of a persistence landscape: (left) from an
   interval to the auxiliary function; (middle) from a barcode to a
   persistence landscape; (right) 3D visualization of the persistence
   landscape.}
   \label{Fig:PersistenceLandscape}
\end{figure}

One of our objectives is to carry out a hypothesis test and make statistical inferences.
For that purpose the barcodes are submitted to a few steps of transformation until we arrive at a test statistic.
We would like to statistically compare two or more different groups of point cloud data.
For example, we would like to carry out a hypothesis test that distinguishes the disk from the annulus.
The usual procedure for such results involves calculating means and variances.
Unfortunately it is not at all clear how to do this for barcodes.
For example, two barcodes need not have a unique Fr\'echet mean.
One solution is to transform the barcode into the \emph{persistence  landscape}, a functional summary of the peristence module introduced by \citet{Bubenik2015}.
The persistence landscape consists of a sequence of functions $\lambda_k:\mathbb{R} \to \mathbb{R}$, where $k=1,2,3,\ldots$.
Here we define these functions via an auxiliary function.
Given $(a,b)$, where $a\leq b$, let $f_{(a,b )}: \real \rightarrow \real$ be the function given by $f_{(a,b)}(t)=\min(t-a, b-t)_+$, where $x_+ =\max
(x,0)$.
Let ${B}$ be a barcode consisting of $m$ intervals with endpoints $\{(a_i, b_i)\}_{i=1}^m$, where $a_i < b_i$.
\begin{definition}
\label{Def:PersLand}
The persistence landscape corresponding to a barcode ${B}$ is the set of functions $\{{\lambda}_k(t): \real \rightarrow \real \}_{k \in \n}$,
where ${\lambda}_k(t)$ is the $k^{th}$ largest value of $\{f_{(a_i,b_i)}(t)\}_{i=1}^m$, and ${\lambda}_k(t)=0$, whenever $k>m$.
\end{definition}
\noindent
These functions may be assembled to give a function $\lambda(k, t)$ defined on $\n\times \real$, which in turn can be extended to $\real^2$ by setting $\lambda(x,t)= \lambda(\lceil x \rceil,t)$ if $x>0$ and $\lambda(x,t)= 0$ otherwise, where $\lceil x\rceil$ denotes the smallest integer obtained when rounding up a real number $x$, and $\n = \left\{ {1,2,3, \ldots } \right\}$ is the set of natural numbers, see Figure~\ref{Fig:PersistenceLandscape}.


\noindent Persistence landscapes for the disk and annulus example are shown in Figure~\ref{Fig:FlowDiskAnnulus}.
Furthermore, we can measure the distance between persistence landscapes as the $p$-norm of their difference.
\begin{definition}
\label{Def:PLdistance}
Let $\lambda \left( {k,t} \right)$ and $\lambda '\left( {k,t} \right)$
be two persistence landscapes. The $p$-Landscape distance between
$\lambda$ and $\lambda '$ is defined by ${\Lambda
  _p}(\lambda,\lambda') = {\left\| {\lambda - \lambda '}
  \right\|_p}$. That is,
     \begin{equation}
        \label{Eq:LandscapeDistance}
        {\Lambda _p}\left( {\lambda ,\lambda '} \right) = {\left[ {\sum\limits_k {\int\limits_{\real} {{{\left| {{\lambda _k}\left( t \right) - {{\lambda '}_k}\left( t \right)} \right|}^p} {dt}} } } \right]^{{1 \mathord{\left/{\vphantom {1 p}} \right. \kern-\nulldelimiterspace} p}}}.
    \end{equation}
\end{definition}

Let $\mathcal{B}$ denote the set of all barcodes, or equivalently, the set of all persistence diagrams. For $p=2$, we can view the persistence landscape as a feature map $\lambda: \mathcal{B} \to L^2(\mathbb{N}\times\mathbb{R})$ to the Hilbert space $L^2(\mathbb{N}\times\mathbb{R})$. From this we obtain a (positive definite) kernel $k:\mathcal{B}\times\mathcal{B}\to\mathbb{R}$ defined by $k(B,B') = \langle \lambda, \lambda' \rangle_{L^2(\mathbb{N}\times\mathbb{R})}$, where $\lambda$ and $\lambda'$ are the persistence landscapes of $B$ and $B'$.
This kernel induces a pseudometric on $\mathcal{B}$ given by $d_k(B,B') = [k(B,B)+k(B',B')-2k(B,B')]^{\frac{1}{2}} = \left\| \lambda-\lambda' \right\|_2 = \Lambda_2(\lambda,\lambda')$.

Now we can establish the main tools needed for our statistical analysis.
Assuming that our persistence landscapes are $p$-integrable, we work in the separable Banach space ${L^p}(\n\times \real)$.
Together with a probability measure on the Borel $\sigma$-algebra, we have also a probability space (e.g. see \citet{Ledoux2002}).
In this space, for any continuous linear functional $f$, the random variable ${f}(\lambda(k,t))$ satisfies the Strong Law of Large Numbers (SLLN) and the Central Limit Theorem (CLT).
In cases where $\lambda$ has finite support we can let $f$ be given by the integration of $\lambda$ multiplied by the indicator function on this support.
Hence we can define a new variable,
\begin{equation}
     X = f\left( {\lambda (k,t)} \right) = \sum\limits_k {\int\limits_{\real} {{\lambda (k,t)}\left(t \right)dt}},
    \label{Eq:RandomVariable}
\end{equation}
whose value corresponds to the total area encompassed by all contours ${\lambda (k,t)}$, $k \in \n$ of a persistence landscape. Since both SLLN and CLT hold, provided a sufficiently large sample the random variable $X$ follows an approximately normal distribution.
This result allows applications of classical statistical methods to point cloud data whose underlying space might be high dimensional or nonlinear.

We conclude the section by setting up a hypothesis test and a corresponding $p$-value based on a permutation test.
A nonparametric test is used due to the low number of samples we have, otherwise a Student's $t$-test would be the preferable choice.
Suppose we wish to compare two {groups} of data, obtained by taking $n_1$ and $n_2$ samples from two geometrical objects.
Let $\lambda_{11} (k,t), \ldots \lambda_{1{n_1}}(k,t)$ and $\lambda_{21} (k,t), \ldots \lambda_{2n_2}(k,t)$ denote the associated persistence landscapes for
a homology in some fixed degree.
Let ${x_{11}}, \ldots ,{x_{1{n_1}}}$ and ${x_{21}}, \ldots ,{x_{2{n_1}}}$ be the associated sample values of random variables $X_1 = f(\lambda_1 (k,t))$ and $X_2 = f(\lambda_2 (k,t))$, respectively, where $f$ is the functional from Equation~\eqref{Eq:RandomVariable}. 
If $\mu_1 $ and $\mu_2 $ are the population means of the random variable $X = f(\lambda (k,t))$ for the two objects, the statistical hypotheses of interest are:
\begin{eqnarray}
\label{Eq:TestHypothesis}
    {H_o}: \mu_1 = \mu_2
    \hskip3mm   \mbox{vs.} \hskip3mm
    {H_a}: \mu_1 \neq \mu_2.
\end{eqnarray}
To test the null-hypothesis we use a two-sample permutation test with statistic,
\begin{equation}
\label{Eq:TestStatistic}
    t = \frac{{\left| {{{\overline X }_1} - {{\overline X }_2}} \right|}}{{\sqrt {\frac{{Var\left( {{X_1}} \right)}}{{{n_1}}} + \frac{{Var\left( {{X_2}} \right)}}{{{n_2}}}} }} \hskip1mm.
\end{equation}
Using the above formula we generate the null distribution as the set of all possible values $t_1, \ldots, t_m$ of the test statistic, calculated for permutations $i=1,\ldots, m$.
Let the observed value of the test statistic be denoted by $t_{obs}$.
Then the $p$-value is obtained as the averaged number of times when the test statistic is at least as extreme as its observed counterpart, $t_q\geq t_{obs}$, where $q \in \left\{ {1, \ldots ,m} \right\}$.

Returning to our disk and annulus example (Figure~\ref{Fig:FlowDiskAnnulus}), in degree 1, a two-sample exact permutation test on this example produces a $p$-value of $0.0028$.
This is as expected, since the disk and the annulus differ in their degree-1 homology because the annulus contains a cycle that is not the boundary of a disk contained in the annulus.
In addition, the p-value in degree 0 is 0.0265. This is somewhat surprising since the disk and the annulus have the same degree-0 homology. We see here that persistent homology is sensitive to geometric differences; the rate at which the points connect differs in the two corresponding filtered simplicial complexes.

In the next section, for our protein data the $p$-value for each degree of homology is obtained from a null distribution of size 1716 (given the nonnegativity of the test statistic).
We apply persistence landscapes to compare closed and open conformations of the maltose-binding protein.
For more on the theory of persistent homology, refer to \citet{Edels2002}, \citet{ZomCar2005}, or \citet{BubenikScott2012}.
For researchers in applied fields, supplementary material in \citet{HGK2012} provides a quick review of persistent homology with hands-on calculations.

%
%

\section{RESULTS AND DISCUSSION}
\label{Sec:ResultsAndDiscussion}

To investigate if a statistically significant difference between the closed and the open conformation can be determined using topological methods, we construct a filtered Vietoris-Rips complex whose persistent homology is calculated.
From persistence intervals we generate persistence landscapes that are further transformed to yield a random variable suitable for hypothesis testing.
The topological approach reinstates the separation between the two conformations, in agreement with initial physical modeling \citep{MBP:DynamicalModel}.
We also demonstrate other ways of distinguishing between open and closed MBP conformations.

\subsection{Snapshots from Evolution}


\begin{figure}[!htbp]\small
   \begin{center}
   \resizebox{!}{0.85\textheight}{
   \includegraphics[angle=0.]{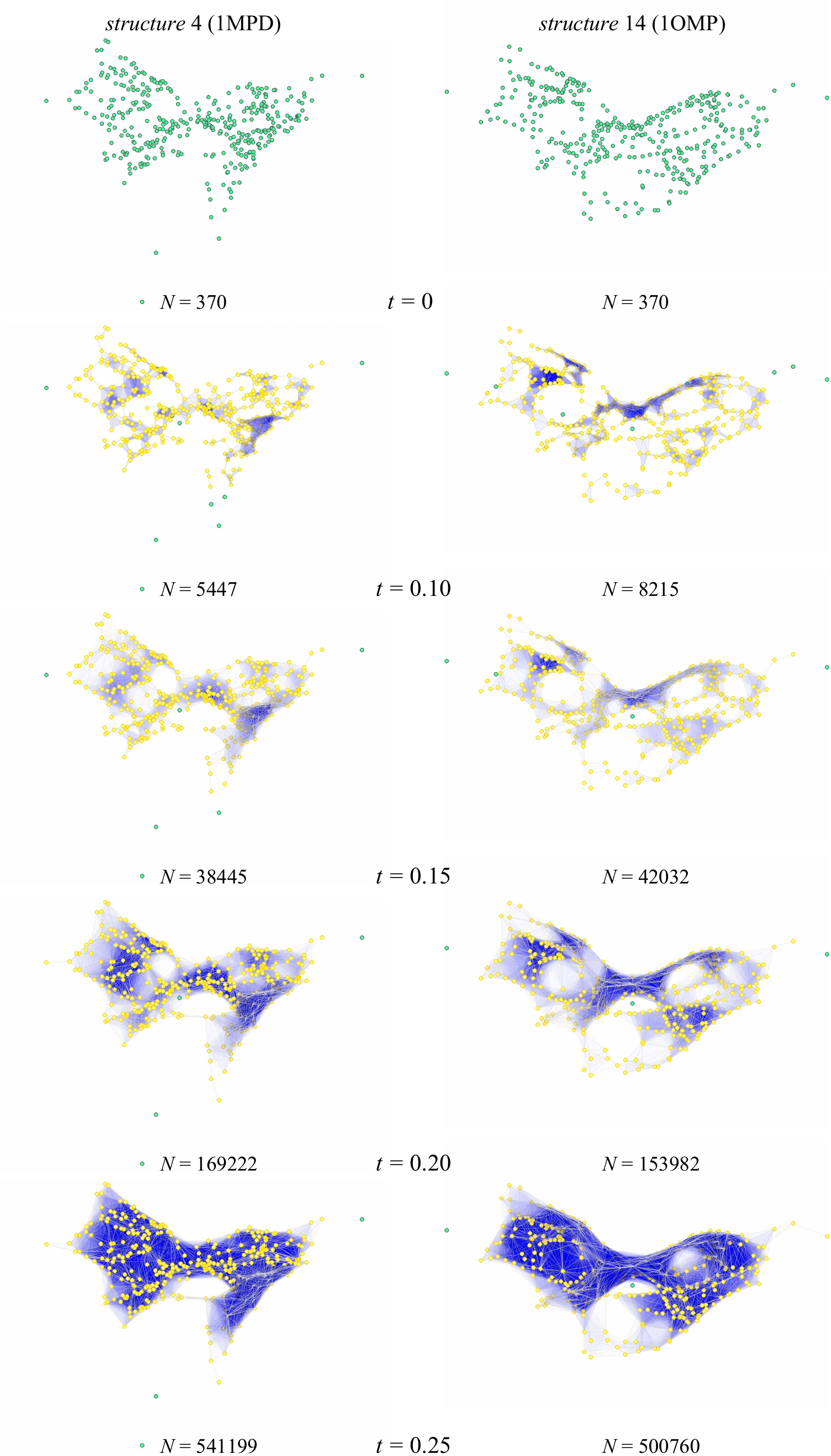}
}
   \end{center}
   \caption{Five snapshots capture the evolution of the filtered Vietoris-Rips complex on the closed-holo 1MPD (left) and the open-apo 1OMP (right) structure of the maltose-binding protein. The complex is constructed on 370 vertices (green circles). The number of vertices that enter the complex (yellow circles) rapidly increases with filtration values. $N$ counts the total number of simplices.}
   \label{Fig:MBP_Evolution}
\end{figure}

Figure~\ref{Fig:MBP_Evolution} portrays a few snapshots from the evolution of the filtered Vietoris-Rips complex constructed on \textsc{Isomap}~\citep{TenenbaumISOMAP} embedded 3D dynamical coordinates of the closed-holo 1MPD and the open-apo 1OMP structures from Table~\ref{Tab:MBPstructures}.
Observe the rapid increase in the number of simplices at higher filtration values.
At filtration $t = 0.3$ the total count of simplices in each structure is about 1.5 million (image is not shown due to excessive memory intake).

\subsection{Visual Comparisons}
\begin{figure}[!htbp]
   \begin{center}
   \resizebox{0.95\columnwidth}{!}{
        \includegraphics[angle=0.,width=0.95\textwidth]{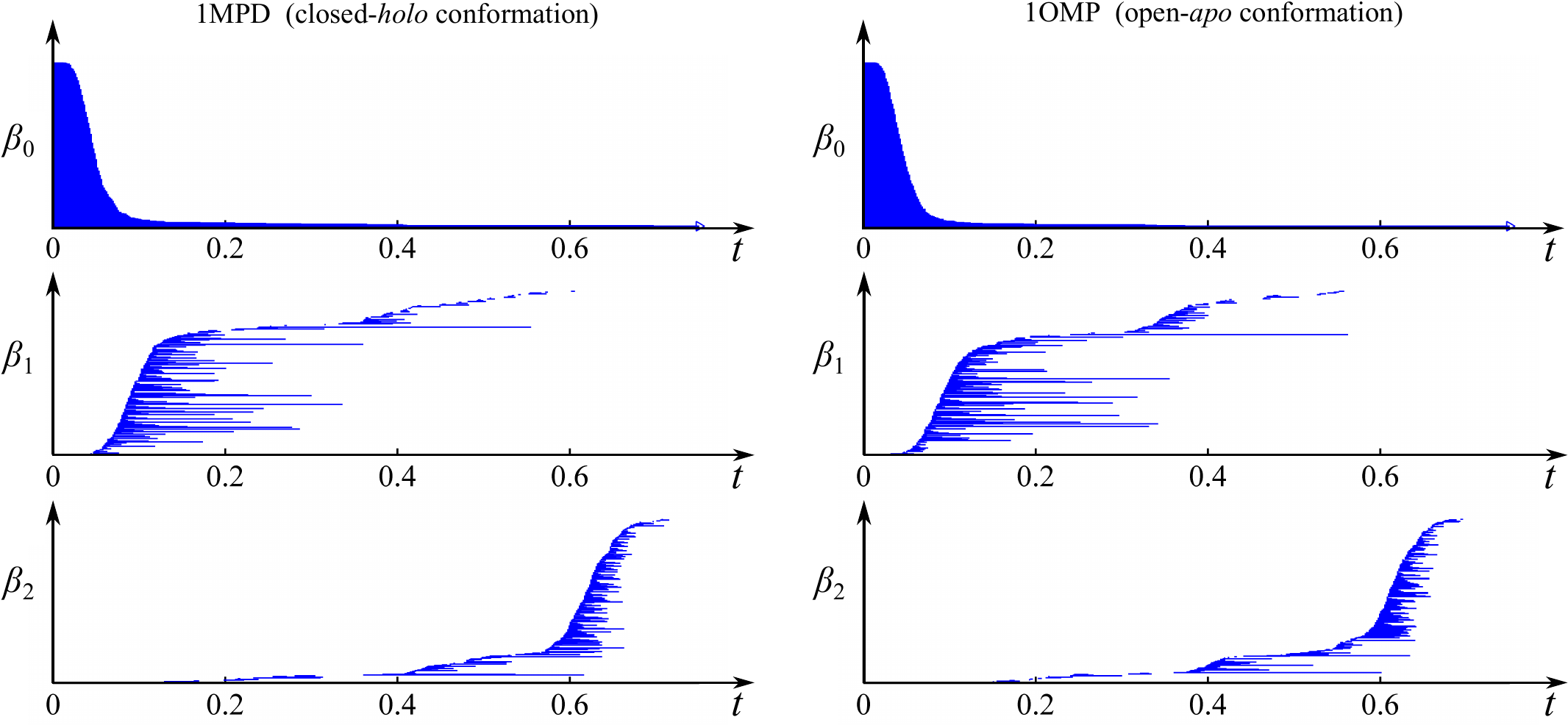}
           }
   \end{center}
   \caption{Barcode plots for the 1MPD closed-holo and the 1OMP open-apo structure corresponding to filtered Vietoris-Rips complexes constructed from the dynamical distance matrices.}
   \label{Fig:MBP_HomologyPlots_fromDistances}
\end{figure}

First, we visually compare barcode plots of the closed and the open MBP conformation. Typical barcodes are shown in Figure~\ref{Fig:MBP_HomologyPlots_fromDistances}.


\begin{figure}[!htbp]
   \begin{center}
   \resizebox{\columnwidth}{!}{
          \includegraphics[angle=0.,width=0.8\textwidth]{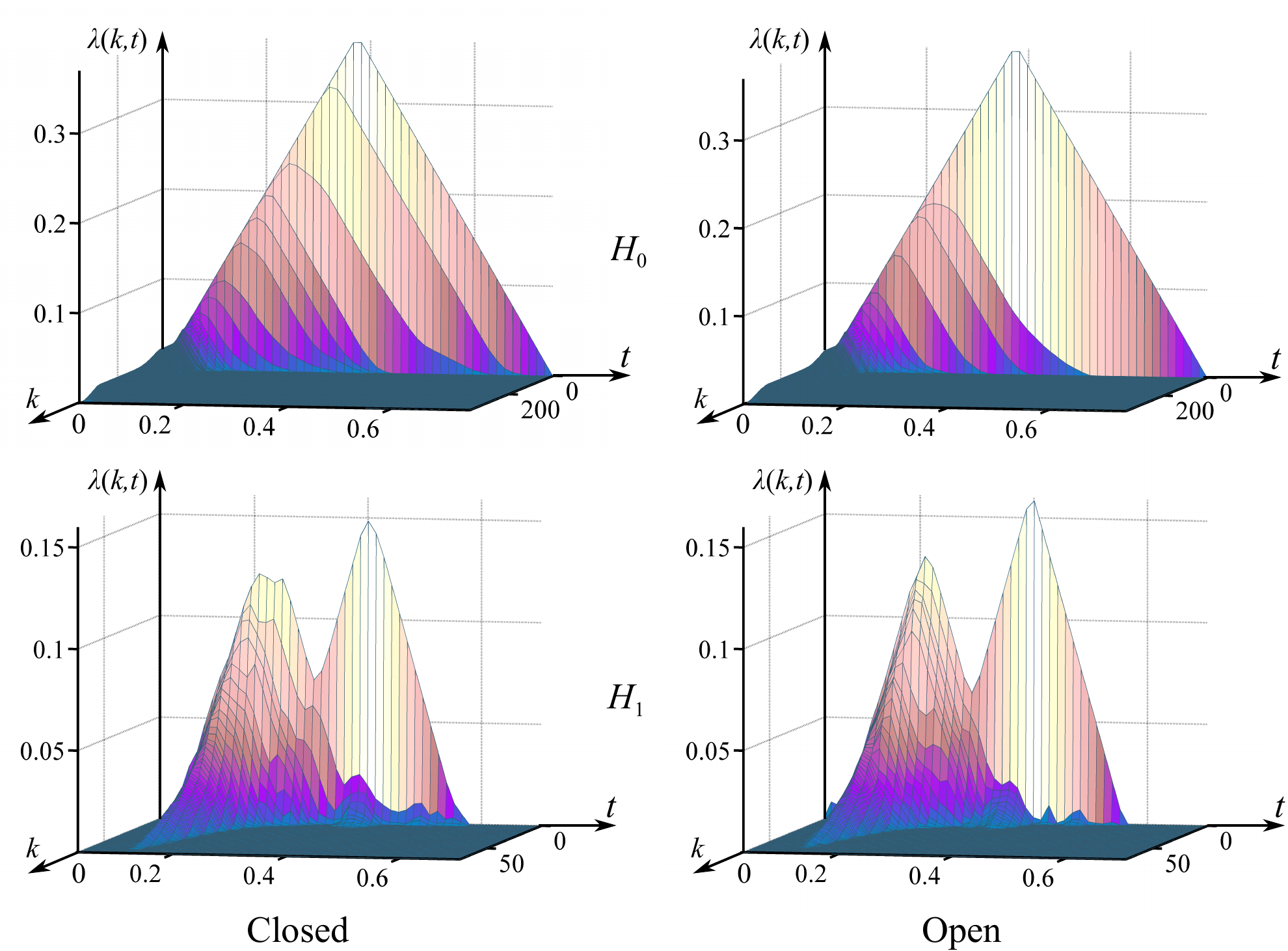}
           }
   \end{center}
   \caption{Mean persistence landscapes of the closed (left) and open (right) MBP structures for degree 0 and degree 1 homology groups.
}
\label{Fig:MBP_AveragePersistenceLandscapes_fromDistances}
\end{figure}

Overall, there is little difference.
In all structures, the degree 1 barcode features one very long and pronounced bar, which is born around time 0.2 and dies shortly before time 0.6.
This bar is represented by a cycle in the Vietoris-Rips complex.
The importance of this most persistent loop will be discussed later.
Unlike the barcodes, we can average the corresponding persistence landscapes and compare the mean persistence landscapes of the closed and open conformations
in Figure~\ref{Fig:MBP_AveragePersistenceLandscapes_fromDistances}.


For dimension 0, the mean shows a greater number of high peaks, implying a greater number of long-persisting components in the dynamical space.
These correspond to the outliers in Figure~\ref{Fig:MBP_Evolution}.
For dimension 1, the two means have a similar layout, both featuring one separate high peak and a cluster of lower peaks; the high distinct peak corresponds to the longest persisting loop.
A small difference appears in the cluster of peaks, as their tops seem less pointed in the case of the closed conformation.

The average persistence landscape in Figure~\ref{Fig:MBP_AveragePersistenceLandscapes_fromDistances} suggests the possibility of systematic differences between the persistent homology of closed and open MBP conformations.
We will next try to see whether persistence landscapes can capture such a difference among the 14 conformations.
We apply support vector machine~(SVM) techniques to persistence landscapes in different ways which we now describe.
In degrees 0, 1 and 2, the persistence landscape consists of functions $\lambda_{1}(t)$, $\lambda_{2}(t)$, $\ldots$, $\lambda_{k}(t)$, with $k=370$, $73$, and $78$, respectively.
{[A12] In practice, we trace a continuous contour $\lambda_i(t)$  ($i = 1,2, \ldots ,k$) through 50 discrete values $\lambda_i(t_j)$, where $t_{\min} \leq t_j\leq t_{\max}$ are equally spaced.
Hence every MBP conformation is associated with a matrix of size ${370 \times 50}$ (for degree 0), ${73 \times 50}$ (for degree 1), and ${78 \times 50}$ (for degree 2).
}

\begin{figure}[!htbp]
   \begin{center}
   \resizebox{\columnwidth}{!}{
  \includegraphics[angle=0.,width=0.2\textwidth]{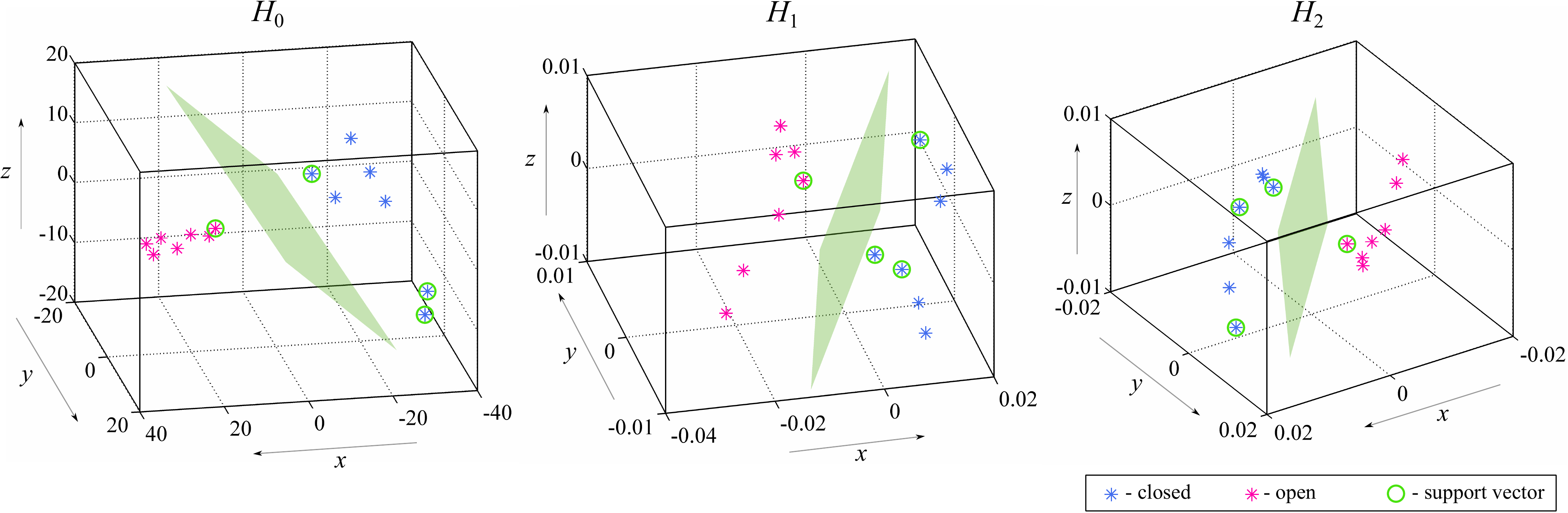}
 }
   \end{center}
\caption{
Results of SVM with linear kernel applied to coordinates obtained from the persistence landscapes of the 14 MBP conformations.
Due to small sample size all data are employed as the training set to yield the hyperplane which demonstrates that the two groups are separable.
Outcome of SVM implemented on the first three principal components of concatenated contours of sample persistence landscapes in homology degrees of 0 (left), 1 (center), and 2 (right). The x, y, z coordinates correspond to the first three principal components.}
\label{Fig:SVM3D_PCA_ISOMAP}
\end{figure}

{
First, the contours $\lambda_1(t)$, $\lambda_2(t)$, $\ldots$, $\lambda_k(t)$ of a persistence landscape are concatenated to form one
long vector in $\real^{1\times50\cdot 370} = \real^{1\times18500}$, $\real^{1\times50\cdot 73} = \real^{1\times3650},$ and $\real^{1\times50\cdot 78} = \real^{1\times3900}$.
Given the 14 samples, we have three feature matrices of sizes $14\times 18500$ (for degree 0), $14\times 3650$ (for degree 1), and $14\times3900$ (for degree 2).
Since the number of variables is high (18500 (degree 0), 3650 (degree 1), and 3900 (degree 2)), relative to the sample size (14), we have over-fitting.
To reduce the dimension we apply the Principal Components Analysis (PCA) to standardized data from each feature matrix (this step is carried out using packages FactoMineR and ade4 of \cite{Rlanguage}).
For each feature matrix the principal components are the eigenvectors of the variance-covariance matrix of the standardized data.
As such, the principal components represent certain linear combinations of the concatenated contours.
These linear combinations correspond to directions with maximum variability and provide a simpler and more parsimonious description of the covariance structure.
The first principal component is the linear combination with  maximum variance, the second principal component is the linear combination with second largest variance and so on.
SVM with a linear kernel \citep{SVM2013} is then performed with the first three principal components, which account for about  $80.42\%$, $52.50\%$, and  $58.68\%$ of the variation with respect to degrees 0, 1, and 2.
Cross-validation is not performed due to small sample size.
The whole data are the training set with the purpose of finding the separating hyperplane between the two groups.
The hyperplane shown in Figure~\ref{Fig:SVM3D_PCA_ISOMAP} illustrates that the two groups are separable.
}


{
Another way of implementing persistence landscapes in statistical analysis uses a $14 \times 14$ matrix of pairwise landscape distances, calculated from the Eq.~(\ref{Eq:LandscapeDistance}) with $p = 2$ ($L_2$-norm).
In such matrix, the $(i, j)$-th entry represents the \emph{$p$-Landscape distance} between the $i$-th and $j$-th conformation ($i,j = 1, 2, \ldots,14$).
This distance matrix serves as input for the Isomap software \citep{TenenbaumISOMAP} which in return provides approximate 3D coordinates of the 14 conformations relative to each other.
The Isomap coordinates are embedded in the metric space induced by the $L_2$ distance.
To asses the error of the Isomap embedding we find the maximum of absolute difference between the landscape $L_2$ distance and the Euclidean distance calculated via Isomap.
The maximum error amounts to 0.043 (deg 0), 0.016 (deg 1), and 0.009 (deg 2).
We also calculate the mean square error; these values are 0.017 (deg 0), 0.007 (deg 1), and 0.004 (deg 2), which is relatively small.
Hence we may proceed with SVM using the 3D Isomap embedded coordinates.
Applying SVM with a linear kernel to the entire dataset we find that the classification boundary accurately separates the two protein groups.
The hyperplanes are  similar to those based on SVM with principal components thus figures are not shown.
}

\subsection{Statistical Inference}
To measure the statistical significance of visually observed differences between the closed and the open conformation we use a permutation test.
For each degree, we calculate fourteen sample values of the random variable $X$ using Eq.~(\ref{Eq:RandomVariable}).
The permutation test carried out at the significance level of $0.05$ yields a $p$-value of $5.83 \times 10^{-4}$ for both homology in
degree 0 and in degree 1.
We obtain the same p-value since in both cases the observed statistic was the most extreme among all 1716 possible permutations.
Hence, at the significance level $\alpha = 0.05$ we have compelling evidence that in the space of dynamical distances the closed and the open MBP conformation significantly differ both in the number of connected components and the number of one dimensional loops.
Concerning the second homology group $H_2$, the test $p$-value of 0.0396 indicates  moderate evidence of difference between closed and open proteins at the level $\alpha = 0.05$.

What can we infer from these results?
In the space of dynamical distances the interpretation of results is not as straightforward as when actual protein coordinates are used.
One may wonder about the meaning of the `number of connected components' in this space; since dynamical distances make the residues with (anti) correlated motion to cluster together, it seems reasonable that results in dimension 0 apply to the number of correlated pieces; similarly, the meaning of the one dimensional `loop' could correspond to `a channel of interaction.'
If so, then we have observed a statistically significant difference between the two conformations in terms of the number of mutually correlated pieces and the number of interaction channels between residues.

In the light of findings from this and the previous section, we note that topological data analysis not only provides different ways to visually compare closed and open MBP conformations, but also gives rise to a hypothesis test for measuring the statistical significance of visually observed differences.
Note that our topological results correspond to those obtained from the initial physical modeling of MBP, see \citet{MBP:DynamicalModel}; this affirms that the first twenty nontrivial normal modes we considered in correlation matrices are sufficient to establish a functional difference between the two protein conformations.

\subsection{Exploring Locations of Residues}

The last part of our research explores locations of residues pertinent for the protein functioning, in particular, active sites and allosteric pathway residues. We also touch upon flexible residues.

\begin{figure}[!htbp]\small
   \begin{center}
   \resizebox{0.65\columnwidth}{!}{
      \includegraphics[angle=0.,width=0.65\textwidth]{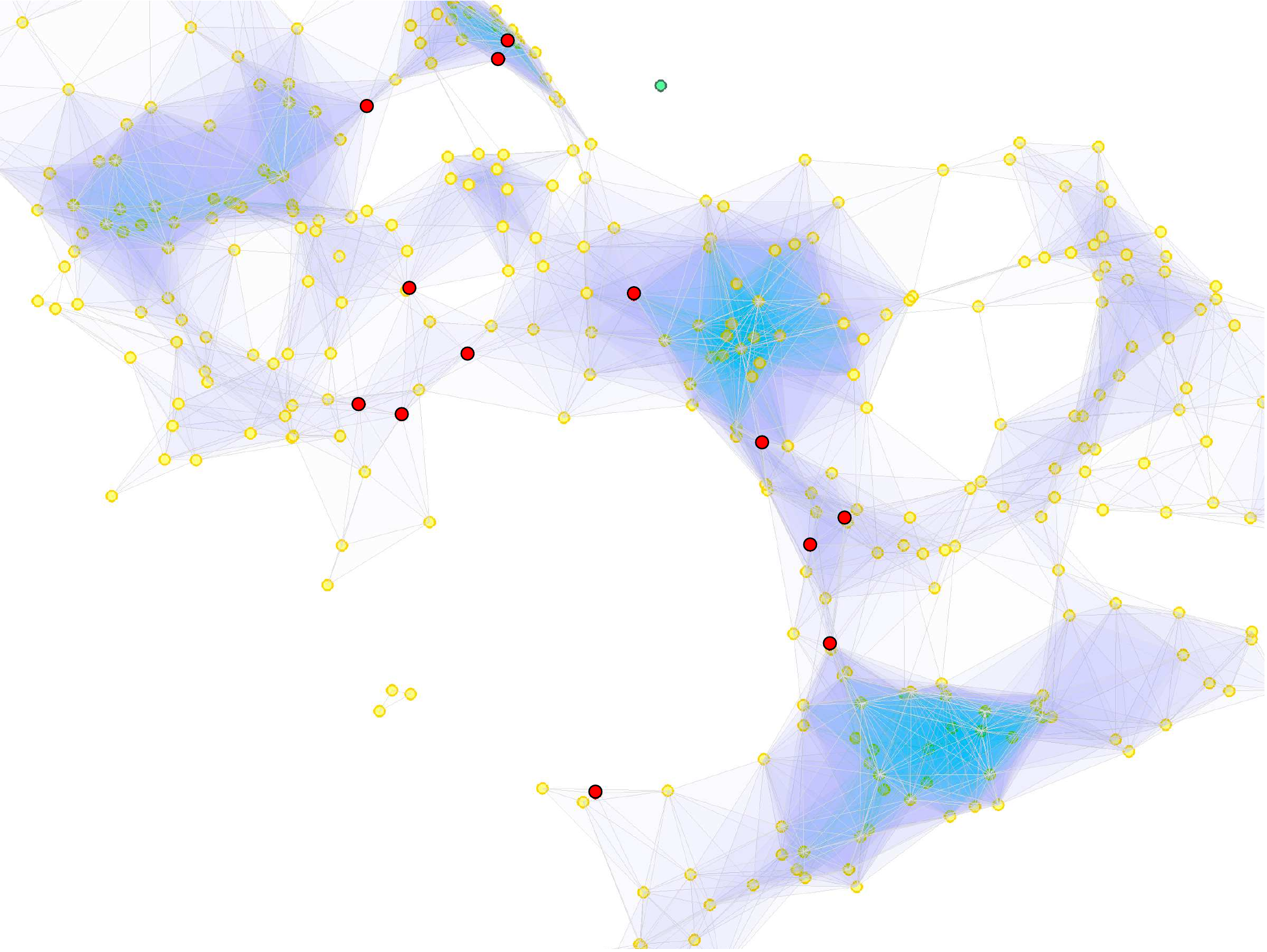}
           }
   \end{center}
   \caption{Active sites (red circles) in the Rips complex of the 1MPD structure, at filtration $t = 0.150$, when the largest loop is still in formation. The majority of active sites lie close to this loop and just a few are positioned around the second largest loop.}
   \label{Fig:1MPD_ActiveSites}
\end{figure}

As we already know, active sites are essential in sugar binding.
They are fairly constrained in their motion inside the protein and well correlated with other residues.
It is thus expected that the dynamical distances from Equation~\eqref{Eq:DistanceMatrix} are rather small for these residues.
We show that \emph{in the dynamical space active sites lie in the vicinity of most persistent loops} in the Vietoris-Rips complex.
This is illustrated for example in the 1MPD structure, see Figure~\ref{Fig:1MPD_ActiveSites}.
Out of thirteen active sites in this structure, ten are positioned near the longest lived loop and the other three dwell in the vicinity of the second
most persistent loop.


A similar result holds for all holo-conformations from Table~\ref{Tab:MBPstructures} -- the bulk of active site residues cluster around the largest loop in the Rips complex and a few are found near other prominent loops.
Therefore the most persistent loops seem to be of special importance.
Next we investigate how the most persistent loop relates to shortest allosteric pathway residues.

\begin{figure}[!htbp]
   \begin{center}
   \resizebox{0.75\columnwidth}{!}{
       \includegraphics[angle=0.,width=0.75\textwidth]{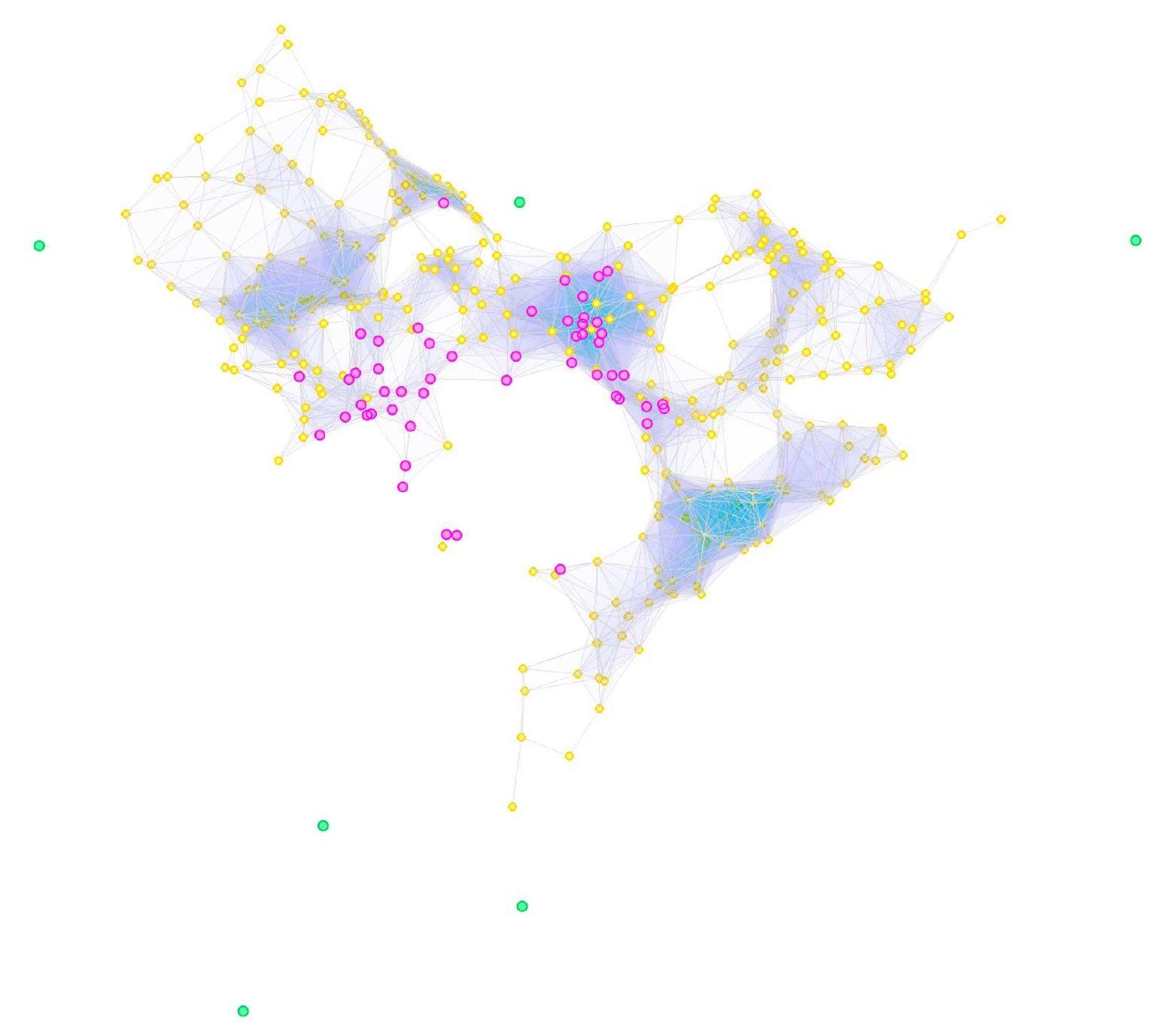}
           }
   \end{center}
   \caption{Layout of shortest allosteric pathway residues (pink dots) of the 1MPD structure inside the Rips complex shown at filtration $t = 150$. There are 51 residues of interest and all but one are scattered around the largest loop. These residues are well correlated with the rest of the protein structure and connect to the Rips complex at early stages of its formations. In contrast, flexible residues (green dots) are among the last ones to connect; they dwell in peripheral regions of the protein where they oscillate with large amplitudes and consequently are the least correlated with other residues.}
   \label{Fig:1MPD_AlloPath_and_FlexibleResidues}
\end{figure}

As mentioned in Section~\ref{Sec:MBP}, allostery provides indirect access to active sites.
The interaction is channeled via a pathway connecting the active site and the allosteric site at the exterior of the protein.
Of interest are the shortest allosteric paths as they are most likely to efficiently transmit stimuli.
Such paths can be computed via the \textsc{AlloPathFinder} software \citep{Tang:2007} which uses Dijkstra's algorithm \citep{Dijkstra:1959}.
Among twenty three residues from the allosteric site in the 1MPD structure \citep{Rizk:2011}, four are best candidates to interact with an
allosteric effector, given the size of their solvent accessible surface area \citep{ASA-View}.
Combined with thirteen active sites we get 52 endpoint assignments, or (since a pair of endpoints can yield multiple solutions) 316 unique shortest allosteric pathways of lengths ranging from five through ten.
After excluding long paths as less potent in conducting impulses from the allosteric effector, we focus on paths of lengths five and six, comprised by 19 and 51 residues, respectively.
The first set is a subset of the other, so there are 51 residues of interest, depicted in Figure~\ref{Fig:1MPD_AlloPath_and_FlexibleResidues}.
Nearly all are located near the most persistent loop in the Rips complex.


Let us now mention flexible residues (see Figure~\ref{Fig:1MPD_AlloPath_and_FlexibleResidues}).
While the majority of vertices connect early on during the evolution of the Rips complex, vertices corresponding to flexible residues are among the last ones to connect.
They strongly oscillate around their equilibrium positions and are poorly correlated with the rest of the protein structure, thus unlikely to take a role in sugar binding.
For more details see \citet{MBP:DynamicalModel}.


\begin{figure}[!htbp]
   \begin{center}
   \resizebox{0.75\columnwidth}{!}{
        \includegraphics[angle=0.,width=0.75\textwidth]{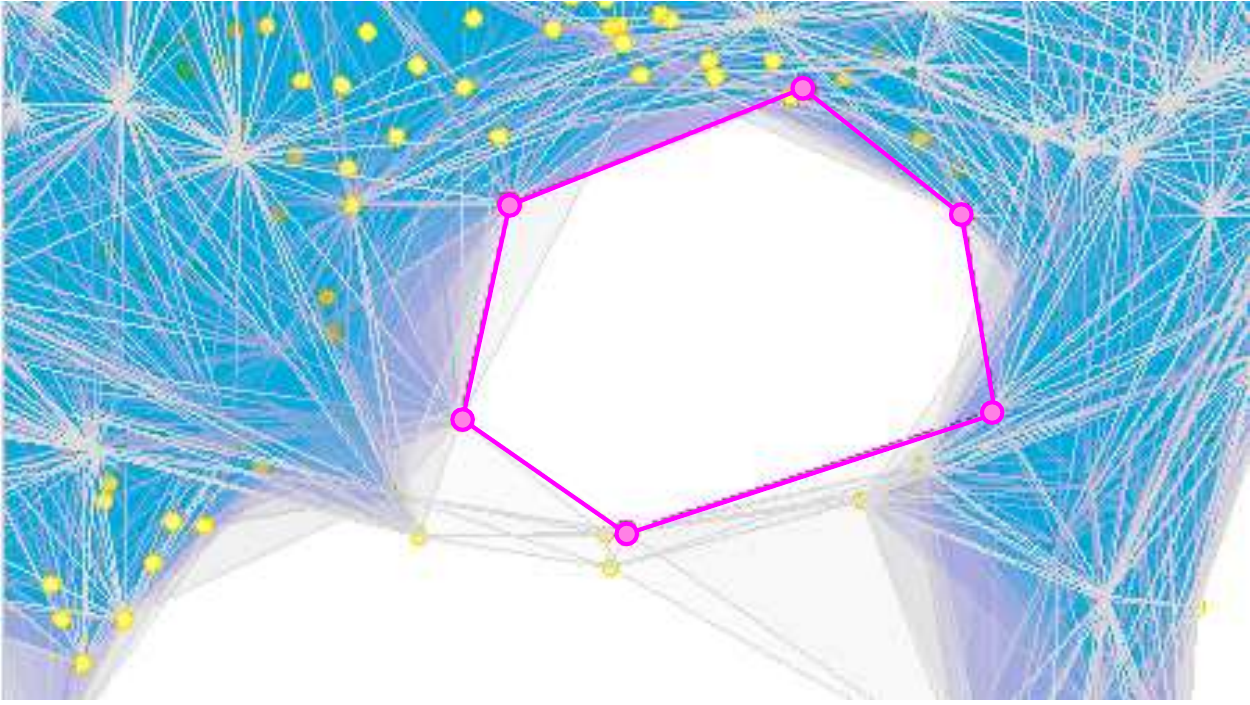}
           }
   \end{center}
   \caption{The short cycle representing the most persistent loop in the 1MPD structure, obtained via the \emph{Short Loop} algorithm of \citet{ShortLoop}. Computations are performed on the Vietoris-Rips complex built on \textsc{Isomap} embedded dynamical coordinates at the filtration parameter which
     corresponds to the midpoint of the lifetime of the most persistent loop ($t = 0.2760$).}
   \label{Fig:1MPD_ShortLoop_HalfTime}
\end{figure}

Results presented so far in this section are qualitative, obtained from a visual representation of the Rips complex.
Now we take a more quantitative approach to the observed most persistent loop.
We use the \emph{Short Loop} software \citep{ShortLoop}.
This algorithm computes the shortest cycle that represents a given homology class of degree one.
To calculate a cycle representing the most persistent loop, for each investigated structure we consider the filtered Rips complex at
filtration value which is the midpoint of the longest interval in the degree-one barcode.
For the resulting cycle for the 1MPD structure, see Figure~\ref{Fig:1MPD_ShortLoop_HalfTime}.
Three out of the six vertices in the shortest cycle belong to the set of 51 interesting allosteric pathway residues; if paths of length seven are included, then four vertices from the short loop belong to the set of allosteric pathway residues.
Moreover, if all twenty three residues from the allosteric site are considered, then five out of six vertices from the short loop belong to the set of allosteric pathway residues, taking into account paths of length up to seven.
Last but not least, we observe that all open conformations feature several short loops, while in all closed conformations the algorithm finds a single short loop (except in the case of the 1FQC structure where an additional smaller cycle appears).

In summary, the most persistent loop in the filtered Rips complex of the maltose-binding protein seems to hold a special biological importance; in all holo-structures the majority of active sites as well as residues that comprise shortest allosteric paths are identified around the most persistent loop in the complex.
Hence the topological approach provides a valuable input in identification of active sites and allosteric pathway residues.
Such information can be useful in future research to single out the best candidates for ligand binding, e.g. in the design of glucose biosensors \citep{Marvin:1997}.
Instead of looking at large number of possible residues, we can focus our attention on those that are in the vicinity of the largest loops, saving time and resources while investigating new protein structures.

%
%

\section{SUMMARY AND FURTHER GOALS}
\label{Sec:Summary}

We have studied a new functional summary for the `shape' of data, the persistence landscape, which was introduced by \citet{Bubenik2015}.
Unlike other topological summaries, e.g. the barcode and the persistence diagram, one can obtain the Fr\'{e}chet mean and Fr\'{e}chet standard deviation of persistence landscapes.
Consequently, persistence landscapes are advantageous for statistical inference.
Following a successful application of this theory to synthetic data from geometrical objects (disk and annulus), we analyzed data sets of
biological importance, namely, fourteen structures of the maltose-binding protein found in the Escherichia coli bacterium.
For that purpose we used dynamical distances obtained from pairwise correlations among amino acid residues {[A15] . The }
correlation matrices originated from the elastic network model developed by taking into account the first twenty non trivial modes of oscillation.

After performing the topological data analysis we confirmed a statistically significant difference between closed and open conformations of the maltose-binding protein, i.e. we were able to discriminate among structural changes pertinent to protein functioning.
SVM with linear kernel showed a good distinction between closed and open protein conformations.
In addition, snapshots of the filtered Vietoris-Rips complex revealed that the most persistent loops host amino acid residues that are actively involved
in sugar uptake.
Moreover, we observed that residues which comprise shortest allosteric interaction pathways also cluster along the largest loop in the complex.
Therefore, the presented topological approach can provide a preliminary screening method in identification of residues susceptible to ligand binding and allosteric manipulation, which could have a potential use in biosensors.
Our confidence is reinforced by the fact that the topological results correspond to the results attained via a physical approach.

For point cloud data of smaller sizes it is natural to apply shape analysis developed by several researchers, \citet{Dryden1998, BP03, BBP09}, and \citet{Bha08}, for example.
However for the maltose-binding protein it is important to analyze the dynamical distance matrices which contain information on the mutual interaction between residues, and shape analysis is not applicable to these correlation matrices.

We conclude with possible future research goals.
\citet{mileyko2011} and \citet{Turner2013} have laid out a theoretical foundation for distributions of persistence diagrams.
They also provided an algorithm for computing the sample Fr\'{e}chet mean of persistence diagrams.
It would be interesting to see how their approach works with our maltose-binding protein data.
\citet{Bendich2011} observed that computing homology by growing Euclidean balls is sensitive to outliers and recommended the use of a metric
derived from a random walk function.
\citet{FasyEtAl2014} investigated analysis of PCD and kernel density estimator and showed that persistent homology based on kernel density estimations is less
sensitive to outliers.
Motivated by works of \citet{Bendich2011} and \citet{FasyEtAl2014} we are currently investigating extensions of this work.
In this article we did not explore the maltose-binding protein using persistence landscapes for homology in degree two.
Analyzing persistence diagrams for degrees 1 and 2, \citet{Gameiro:2012} were able to successfully predict the compressibility of various protein
structures (\citet{Gekko:1986}).
It would be interesting to study the compressibility of protein structures based on higher degree persistence landscapes.

%
%
\section*{ACKNOWLEDGMENTS}

{\footnotesize
V.K.N. would like to thank to Walter Roberson from the Matlab Team for a clarification on SVM.
\newline
\noindent
P.~B. gratefully acknowledges support from AFOSR grant \# FA9550-13-1-0115.
\newline
\noindent
G.~H. acknowledges funding support provided by the University of Alberta Orthodontic Division's McIntyre Memorial Fund and NSERC Discovery Grant 293180.
\newline
\noindent
Computations in this research were largely enabled by resources provided by WestGrid and Compute/Calcul Canada.
}

%
%
\section*{GRAPHICS SOFTWARE}
The following software was used to generate and process images.

{\footnotesize
\begin{description}

\item[\textbf{Figure~\ref{Fig:FlowDiskAnnulus}:}]
\noindent
 Evolution plots (a) were generated using the \textsc{PLEX} library \citep{PLEX} called in \citet{MATLAB2005} and further edited in \citet{MATLAB2011}, which was also the the source of other plots. Barcode plots (b) were made via \textsc{javaPlex} library \citep{javaPlex}. Persistence landscapes (c) generated using codes provided by \citet{Bubenik2015}. All images were formatted in \citet{Inkscape}.

\item[\textbf{Figure~\ref{Fig:1MPD_CorrelationAndDistanceMatrix}}:]
\noindent
 The correlation matrix was retrieved from the ANM web server \citep{ANM:2006} and based on this the distance matrix was calculated using \citet{MATLAB2011}, which was also used for visualizing both matrices. Images were formatted in \citet{Inkscape}.

\item[\textbf{Figures~\ref{Fig:MBP_ScreePlots}}:]
\noindent
 Plots were created via the \textsc{Isomap} library \citep{TenenbaumISOMAP} called in \citet{MATLAB2011} and formatted using \citet{Inkscape}.
 
\item[\textbf{Figure~\ref{Fig:PersistenceLandscape}},~\textbf{\ref{Fig:MBP_AveragePersistenceLandscapes_fromDistances}}:]
\noindent
 Images were created in \citet{MATLAB2011} and formatted using \citet{Inkscape}. Persistence landscape codes were provided by \citet{Bubenik2015}.

\item[\textbf{Figures~\ref{Fig:MBP_Evolution}}, \textbf{\ref{Fig:1MPD_ActiveSites}--\ref{Fig:1MPD_ShortLoop_HalfTime}}:]
\noindent
 Plots generated via the \textsc{PLEX} library \citep{PLEX} called in \citet{MATLAB2005}, further edited in \citet{MATLAB2011}, and formatted using \citet{Inkscape}.

\item[\textbf{Figure~\ref{Fig:MBP_HomologyPlots_fromDistances}}:]
\noindent
 With the aid of computing resources provided by WestGrid and
 Compute/Calcul Canada, plots were created via the \textsc{javaPlex} library \citep{javaPlex}, called in \citet{MATLAB2011}. Plots were formatted using \citet{Inkscape}.
 
\item[\textbf{Figure~\ref{Fig:SVM3D_PCA_ISOMAP}}]
\noindent
 This image was generated in \citet{MATLAB2011} and formatted using \citet{Inkscape}. Code for making a 3D plot was adopted from \cite{SVM2013} and accordingly modified.

\end{description}
}

%
%
%

\bibliographystyle{apalike2}
\bibliography{MBP_Bibliography}

\begin{thebibliography}{}

\bibitem[{Ahmad} et~al., 2004]{ASA-View}
{Ahmad}, S., {Gromiha}, M., {Fawareh}, H., \& {Sarai}, A. (2004).
\newblock Asa-view: Solvent acesitiblity graphics for proteins.
\newblock {available at \texttt{http://www.abren.net/asaview/}}.

\bibitem[Amitai et~al., 2004]{Amitai:2004}
Amitai, G., Shemesh, A., Sitbon, E., Shklar, M., Netanely, D., Venger, I., \&
  Pietrokovski, S. (2004).
\newblock {Network Analysis of Protein Structures Identifies Functional
  Residues}.
\newblock {\em Journal of Molecular Biology}, 344(4), 1135--1146.

\bibitem[Atilgan et~al., 2001]{Atilgan:2001}
Atilgan, A.~R., Durell, S.~R., Jernigan, R.~L., Demirel, M.~C., Keskin, O., \&
  Bahar, I. (2001).
\newblock {Anisotropy of Fluctuation Dynamics of Proteins with an Elastic
  Network Model}.
\newblock {\em Biophysical Journal}, 80(1), 505--515.

\bibitem[{Bandulasiri} et~al., 2009]{BBP09}
{Bandulasiri}, A., {Bhattacharya}, R.~N., \& {Patrangenaru}, V. (2009).
\newblock {Nonparametric Inference for Extrinsic Means on
  Size-and-(Reflection)-Shape Manifolds with Applications in Medical Imaging}.
\newblock {\em Journal of Multivariate Analysis}, 100, 1867--1882.

\bibitem[{Bendich} et~al., 2011]{Bendich2011}
{Bendich}, P., {Galkovskyi}, T., \& {Harer}, J. (2011).
\newblock Improving homology estimates with random walks.
\newblock {\em Inverse Problems}, 27, 124002.

\bibitem[{Bernstein} et~al., 1977]{PDB}
{Bernstein}, F.~C., {Koetzle}, T.~F., {Williams}, G.~J., {Meyer Jr.}, E.~E.,
  {Brice}, M.~D., {Rodgers}, J.~R., {Kennard}, O., {Shimanouchi}, T., \&
  {Tasumi}, M. (1977).
\newblock {The Protein Data Bank: A Computer-based Archival File For
  Macromolecular Structures}.
\newblock {\em J. Mol. Biol.}, 112, 535.

\bibitem[{Bhattacharya}, 2008]{Bha08}
{Bhattacharya}, A. (2008).
\newblock {Statistical Analysis on Manifolds: A Nonparametric Approach for
  Inference on Shape Spaces}.
\newblock {\em Sankhya, the Indian Journal of Statistics}, 70-A, 223--266.

\bibitem[{Bhattacharya} \& {Patrangenaru}, 2003]{BP03}
{Bhattacharya}, R. \& {Patrangenaru}, V. (2003).
\newblock {Large sample theory of intrinsic and extrinsic sample means on
  manifolds~I}.
\newblock {\em Annals of Statistics}, 31, 1--29.

\bibitem[{Boos} \& {Shuman}, 1998]{BoosShuman:1998}
{Boos}, W. \& {Shuman}, H. (1998).
\newblock {Maltose{/}Maltodextrin System of Escherichia coli: Transport,
  Metabolism, and Regulation}.
\newblock {\em Microbiology and Molecular Biology Reviews}, 62(1).

\bibitem[Bradley et~al., 2008]{Bradley:2008}
Bradley, M.~J., Chivers, P.~T., \& Baker, N.~A. (2008).
\newblock {Molecular Dynamics Simulation of the Escherichia coli NikR Protein:
  Equilibrium Conformational Fluctuations Reveal Interdomain Allosteric
  Communication Pathways}.
\newblock {\em Journal of Molecular Biology}, 378(5), 1155--1173.

\bibitem[{Bubenik}, 2015]{Bubenik2015}
{Bubenik}, P. (2015).
\newblock Statistical topological data analysis using persistence landscapes.
\newblock {\em Journal of Machine Learning Research}, 16, 77--102.

\bibitem[Bubenik et~al., 2010]{Bubenik2010}
Bubenik, P., Carlsson, G., Kim, P.~T., \& Luo, Z.-M. (2010).
\newblock Statistical topology via {M}orse theory persistence and nonparametric
  estimation.
\newblock In {\em Algebraic methods in statistics and probability {II}}, volume
  516 of {\em Contemp. Math.}  (pp.\ 75--92). Providence, RI: Amer. Math. Soc.

\bibitem[{Bubenik} \& {Scott}, 2014]{BubenikScott2012}
{Bubenik}, P. \& {Scott}, J. (2014).
\newblock Categorification of persistent homology.
\newblock {\em Discrete \& Computational Geometry}, 51(3), 600--627.

\bibitem[Cavasotto et~al., 2005]{Cavasotto:2005}
Cavasotto, C.~N., Kovacs, J.~A., \& Abagyan, R.~A. (2005).
\newblock {Representing Receptor Flexibility in Ligand Docking through Relevant
  Normal Modes}.
\newblock {\em Journal of the American Chemical Society}, 127(26), 9632--9640.

\bibitem[{Chazal} et~al., 2014]{ChazalEtAl2014}
{Chazal}, F., {Fasy}, B., {Lecci}, F., {Rinaldo}, A., {Singh}, A., \&
  {Wasserman}, L. (2014).
\newblock {On the Bootstrap for Persistence Diagrams and Landscapes}.
\newblock {\em Modeling and Analysis of Information Systems}, 20(6), 96--105.

\bibitem[Chazal et~al., 2014]{ChazalEtAl2013}
Chazal, F., Fasy, B.~T., Lecci, F., Rinaldo, A., \& Wasserman, L. (2014).
\newblock {Stochastic Convergence of Persistence Landscapes and Silhouettes}.
\newblock In {\em Proceedings of the Thirtieth Annual Symposium on
  Computational Geometry}, SOCG'14  (pp.\ 474--483).  New York, NY, USA: ACM.

\bibitem[{Collins} et~al., 2004]{CZCG2004a}
{Collins}, A., {Zomorodian}, A., {Carlsson}, G., \& {Guibas}, L.~J. (2004).
\newblock A barcode shape descriptor for curve point cloud data.
\newblock {\em Computers and Graphics}, 28, 881--894.

\bibitem[{de Silva} \& {Perry}, 2005]{PLEX}
{de Silva}, V. \& {Perry}, P. (2005).
\newblock {Plex: A MATLAB library for studying simplicial homology}.
\newblock available at \texttt{http://comptop.stanford.edu/programs/plex}.

\bibitem[Dijkstra, 1959]{Dijkstra:1959}
Dijkstra, E. (1959).
\newblock A note on two problems in connexion with graphs.
\newblock {\em Numerische Mathematik}, 1(1), 269--271.

\bibitem[{Dryden} \& {Mardia}, 1998]{Dryden1998}
{Dryden}, I.~L. \& {Mardia}, K.~V. (1998).
\newblock {\em {Statistical Shape Analysis}}.
\newblock New York: John Wiley and Sons.

\bibitem[Duan et~al., 2001]{Duan:2001}
Duan, X., Hall, J.~A., Nikaido, H., \& Quiocho, F.~A. (2001).
\newblock {Crystal structures of the maltodextrin/maltose-binding protein
  complexed with reduced oligosaccharides: flexibility of tertiary structure
  and ligand binding}.
\newblock {\em Journal of Molecular Biology}, 306(5), 1115--1126.

\bibitem[Duan \& Quiocho, 2002]{Duan:2002}
Duan, X. \& Quiocho, F.~A. (2002).
\newblock {Structural Evidence for a Dominant Role of Nonpolar Interactions in
  the Binding of a Transport/Chemosensory Receptor to Its Highly Polar
  Ligands}.
\newblock {\em Biochemistry}, 41(3), 706--712.

\bibitem[{Edelsbrunner} \& {Harer}, 2010]{EHbook2010}
{Edelsbrunner}, H. \& {Harer}, J. (2010).
\newblock {\em {Computational Topology An Introduction}}.
\newblock Providence Rhode Island: American Mathematical Society.

\bibitem[{Edelsbrunner} et~al., 2002]{Edels2002}
{Edelsbrunner}, H., {Letscher}, D., \& {Zomorodian}, A. (2002).
\newblock Topological persistence and simplification.
\newblock {\em Discrete and Computational Geometry}, 28, 511--533.

\bibitem[Eyal et~al., 2006]{ANM:2006}
Eyal, E., Yang, L.-W., Bahar, I., \& Tramontano, A. (2006).
\newblock Anisotropic network model: systematic evaluation and a new web
  interface.
\newblock {\em Bioinformatics}, 22, 2619--2627.

\bibitem[{Fasy} et~al., 2014]{FasyEtAl2014}
{Fasy}, B.~T., {Lecci}, F., {Rinaldo}, A., {Wasserman}, L., {Balakrishnan}, S.,
  \& {Singh}, A. (2014).
\newblock Confidence sets for persistence diagrams.
\newblock {\em Ann. Statist.}, 42(6), 2301--2339.

\bibitem[{Gameiro} et~al., 2012]{Gameiro:2012}
{Gameiro}, M., {Hiraoka}, Y., {Izumi}, S., {Kramar}, M., {Mischaikow}, K., \&
  {Nanda}, V. (2012).
\newblock Topological measurement of protein compressibility via persistence
  diagrams.
\newblock In {\em The Global COE Program}, volume~6 of {\em MI Preprint Series}
   (pp.\ 1--10).  Fukuoka, Japan: Math for Industry Education \& Research Hub
  Kyushu University.

\bibitem[Gekko \& Hasegawa, 1986]{Gekko:1986}
Gekko, K. \& Hasegawa, Y. (1986).
\newblock {Compressibility--structure relationship of globular proteins}.
\newblock {\em Biochemistry}, 25(21), 6563--6571.

\bibitem[Gould \& Shilton, 2010]{Gould:2010}
Gould, A.~D. \& Shilton, B.~H. (2010).
\newblock {Studies of the Maltose Transport System Reveal a Mechanism for
  Coupling ATP Hydrolysis to Substrate Translocation without Direct Recognition
  of Substrate}.
\newblock {\em Journal of Biological Chemistry}, 285(15), 11290--11296.

\bibitem[Hatcher, 2002]{Hatcher}
Hatcher, A. (2002).
\newblock {\em Algebraic topology}.
\newblock Cambridge: Cambridge University Press.

\bibitem[{Heo} et~al., 2012]{HGK2012}
{Heo}, G., {Gamble}, J., \& {Kim}, P.~T. (2012).
\newblock Topological analysis of variance and the maxillary complex.
\newblock {\em Journal of the American Statistical Association}, 107, 477--492.

\bibitem[HKF, 2013]{SVM2013}
HKF (2013).
\newblock {How to plot a hyper plane in 3D for the SVM results?}
\newblock \texttt{http://stackoverflow.com/a/19969412}.

\bibitem[{Hudault} et~al., 2001]{Hudault:2001}
{Hudault}, S., {Guignot}, J., \& {Servin}, A.~L. (2001).
\newblock {Escherichia coli strains colonising the gastrointestinal tract
  protect germfree mice against Salmonella typhimurium infection}.
\newblock {\em {Gut}}, 49(1), 47--55.

\bibitem[Inkscape, 2010]{Inkscape}
Inkscape (2010).
\newblock Inkscape: open source vector graphics editor.
\newblock {Free Software Foundation, Inc., available at
  \texttt{http://inkscape.org/}}.

\bibitem[Kasahara et~al., 2014]{Kasahara:2014}
Kasahara, K., Fukuda, I., \& Nakamura, H. (2014).
\newblock {A Novel Approach of Dynamic Cross Correlation Analysis on Molecular
  Dynamics Simulations and Its Application to Ets1 Dimer--DNA Complex}.
\newblock {\em PLoS ONE}, 9(11), e112419.

\bibitem[Kobryn et~al., 2014]{Kobryn:2014}
Kobryn, A.~E., Nikoli\'{c}, D., Lyubimova, O., Gusarov, S., \& Kovalenko, A.
  (2014).
\newblock Dissipative particle dynamics with an effective pair potential from
  integral equation theory of molecular liquids.
\newblock {\em The Journal of Physical Chemistry B}, 118(41), 12034--12049.

\bibitem[{Kovacev-Nikolic}, 2012]{MScThesis}
{Kovacev-Nikolic}, V. (2012).
\newblock {Persistent Homology in Analysis of Point-Cloud Data}.
\newblock Master's thesis, Department of Mathematical and Statistical Sciences,
  University of Alberta.

\bibitem[Ledoux \& Talagrand, 2002]{Ledoux2002}
Ledoux, M. \& Talagrand, M. (2002).
\newblock {\em Probability in Banach Spaces: Isoperimetry and Processes}.
\newblock A Series of Modern Surveys in Mathematics Series. Springer, first
  reprint 2002 edition.

\bibitem[Lockless \& Ranganathan, 1999]{Lockless:1999}
Lockless, S.~W. \& Ranganathan, R. (1999).
\newblock {Evolutionarily Conserved Pathways of Energetic Connectivity in
  Protein Families}.
\newblock {\em Science}, 286(5438), 295--299.

\bibitem[Marvin et~al., 1997]{Marvin:1997}
Marvin, J.~S., Corcoran, E.~E., Hattangadi, N.~A., Zhang, J.~V., Gere, S.~A.,
  \& Hellinga, H.~W. (1997).
\newblock The rational design of allosteric interactions in a monomeric protein
  and its applications to the construction of biosensors.
\newblock {\em Proceedings of the National Academy of Sciences}, 94(9),
  4366--4371.

\bibitem[MATLAB, 2005]{MATLAB2005}
MATLAB (2005).
\newblock Matlab release 14.
\newblock The MathWorks Inc., Natick, Massachusetts, USA.

\bibitem[MATLAB, 2011]{MATLAB2011}
MATLAB (2011).
\newblock Matlab and statistics toolbox release 2011a.
\newblock The MathWorks Inc., Natick, Massachusetts, USA.

\bibitem[{McNaught} \& {Wilkinson}, 1997]{IUPAC}
{McNaught}, A.~D. \& {Wilkinson}, A. (1997).
\newblock {\em {IUPAC Compendium of Chemical Terminology}}.
\newblock Oxford: Blackwell Scientific Publications, 2 edition.

\bibitem[{Mileyko} et~al., 2011]{mileyko2011}
{Mileyko}, Y., {Mukherjee}, S., \& {Harer}, J. (2011).
\newblock Probability measures on the space of persistence diagrams.
\newblock {\em Inverse Problems}, 27, 124007.

\bibitem[Nikoli\'{c} et~al., 2012]{Nikolic:2012}
Nikoli\'{c}, D., Blinov, N., Wishart, D., \& Kovalenko, A. (2012).
\newblock 3d-rism-dock: A new fragment-based drug design protocol.
\newblock {\em Journal of Chemical Theory and Computation}, 8(9), 3356--3372.

\bibitem[Nikoli\'{c} \& {Kovacev-Nikolic}, 2013]{MBP:DynamicalModel}
Nikoli\'{c}, D. \& {Kovacev-Nikolic}, V. (2013).
\newblock Dynamical model of the maltose-binding protein.
\newblock manuscript, Research Gate DOI: {10.13140/2.1.3269.8883}.

\bibitem[Quiocho et~al., 1997]{Quiocho:1997}
Quiocho, F.~A., Spurlino, J.~C., \& Rodseth, L.~E. (1997).
\newblock {Extensive features of tight oligosaccharide binding revealed in
  high-resolution structures of the maltodextrin transport/chemosensory
  receptor}.
\newblock {\em Structure}, 5(8), 997--1015.

\bibitem[{R Development Core Team}, 2008]{Rlanguage}
{R Development Core Team} (2008).
\newblock {\em R: A Language and Environment for Statistical Computing}.
\newblock R Foundation for Statistical Computing, Vienna, Austria.
\newblock {ISBN} 3-900051-07-0.

\bibitem[{Reininghause} et~al., 2015]{Reininghaus2014}
{Reininghause}, J., {Huber}, S., {Bauer}, U., \& {Kwitt}, R. (2015).
\newblock A stable multi-scale kernel for topological machine learning.
\newblock In {\em Proc. 2015 IEEE Conf. Comp. Vision \& Pat. Rec. (CVPR '15)}
  (pp.\ 4741--4748).  Boston, MA, USA.

\bibitem[Rizk et~al., 2011]{Rizk:2011}
Rizk, S.~S., Paduch, M., Heithaus, J.~H., Duguid, E.~M., Sandstrom, A., \&
  Kossiakoff, A.~A. (2011).
\newblock {Allosteric control of ligand-binding affinity using engineered
  conformation-specific effector proteins}.
\newblock {\em Nature Structural \& Molecular Biology}, 18(4), 437--442.

\bibitem[Rubin et~al., 2002]{Rubin:2002}
Rubin, S.~M., Lee, S.-Y., Ruiz, E.~J., Pines, A., \& Wemmer, D.~E. (2002).
\newblock {Detection and Characterization of Xenon-binding Sites in Proteins by
  129Xe NMR Spectroscopy}.
\newblock {\em Journal of Molecular Biology}, 322(2), 425--440.

\bibitem[{Seeliger} \& {de Groot}, 2010]{Seeliger:2010}
{Seeliger}, D. \& {de Groot}, B.~L. (2010).
\newblock {Conformational Transitions upon Ligand Binding: Holo-Structure
  Prediction from Apo Conformations}.
\newblock {\em PLoS Comput. Biol.}, 6(1), e1000634.

\bibitem[Sharff et~al., 1992]{Sharff:1992}
Sharff, A.~J., Rodseth, L.~E., Spurlino, J.~C., \& Quiocho, F.~A. (1992).
\newblock {Crystallographic evidence of a large ligand-induced hinge-twist
  motion between the two domains of the maltodextrin binding protein involved
  in active transport and chemotaxis}.
\newblock {\em Biochemistry}, 31(44), 10657--10663.

\bibitem[Shilton et~al., 1996]{Shilton:1996}
Shilton, B.~H., Shuman, H.~A., \& Mowbray, S.~L. (1996).
\newblock {Crystal Structures and Solution Conformations of a Dominant-negative
  Mutant of Escherichia coli Maltose-binding Protein}.
\newblock {\em Journal of Molecular Biology}, 264(2), 364--376.

\bibitem[{Szmelcman} et~al., 1976]{Szmelcman:1976}
{Szmelcman}, S., {Schwartz}, M., {Silhavy}, T.~J., \& {Boos}, W. (1976).
\newblock {Maltose Transport in Escherichia coli K12}.
\newblock {\em European Journal of Biochemistry}, 65(1), 13--19.

\bibitem[{Tamal} et~al., 2011]{ShortLoop}
{Tamal}, K.~D., {Jian}, S., \& {Yusu}, W. (2011).
\newblock Approximating cycles in a shortest basis of the first homology group
  from point data.
\newblock {\em Inverse Problems}, 27(12), 124004.

\bibitem[Tang et~al., 2007]{Tang:2007}
Tang, S., Liao, J.-C., Dunn, A.~R., Altman, R.~B., Spudich, J.~A., \& Schmidt,
  J.~P. (2007).
\newblock {Predicting Allosteric Communication in Myosin via a Pathway of
  Conserved Residues}.
\newblock {\em Journal of Molecular Biology}, 373(5), 1361--1373.
\newblock available at \texttt{https://simtk.org/home/allopathfinder}.

\bibitem[{Tausz} et~al., 2011]{javaPlex}
{Tausz}, A., {Vejdemo-Johansson}, M., \& {Adams}, H. (2011).
\newblock {JavaPlex: A research software package for persistent (co)homology}.
\newblock {available at \texttt{http://code.google.com/javaplex}}.

\bibitem[{Tenenbaum} et~al., 2000]{TenenbaumISOMAP}
{Tenenbaum}, J.~B., {de Silva}, V., \& {Langford}, J.~C. (2000).
\newblock Isomap: A global geometric framework for nonlinear dimensionality
  reduction.
\newblock {\em Science}, 290(5500), 2319--2323.
\newblock available at \texttt{http://isomap.stanford.edu}.

\bibitem[Tobi \& Bahar, 2005]{Tobi:2005}
Tobi, D. \& Bahar, I. (2005).
\newblock Structural changes involved in protein binding correlate with
  intrinsic motions of proteins in the unbound state.
\newblock {\em Proceedings of the National Academy of Sciences of the United
  States of America}, 102(52), 18908--18913.

\bibitem[Turner et~al., 2014]{Turner2013}
Turner, K., Mileyko, Y., Mukherjee, S., \& Harer, J. (2014).
\newblock Fr\'{e}chet means for distributions of persistence diagrams.
\newblock {\em Discrete \& Computational Geometry}, 52(1), 44--70.

\bibitem[{Van Houdt} \& {Michiels}, 2005]{VanHoudt:2005}
{Van Houdt}, R. \& {Michiels}, C.~W. (2005).
\newblock {Role of bacterial cell surface structures in Escherichia coli
  biofilm formation}.
\newblock {\em Research in Microbiology}, 156(5-6), 626--633.

\bibitem[Xia \& Wei, 2014]{Xia:2014}
Xia, K. \& Wei, G.-W. (2014).
\newblock Persistent homology analysis of protein structure, flexibility, and
  folding.
\newblock {\em International Journal for Numerical Methods in Biomedical
  Engineering}, 30(8), 814--844.

\bibitem[Xia \& Wei, 2015a]{Xia:2015a}
Xia, K. \& Wei, G.-W. (2015a).
\newblock Multidimensional persistence in biomolecular data.
\newblock {\em Journal of Computational Chemistry}, 36(20), 1502--1520.

\bibitem[Xia \& Wei, 2015b]{Xia:2015b}
Xia, K. \& Wei, G.-W. (2015b).
\newblock Persistent topology for cryo-em data analysis.
\newblock {\em International Journal for Numerical Methods in Biomedical
  Engineering}, (pp.\ n/a--n/a).

\bibitem[{Zomorodian} \& {Carlsson}, 2005]{ZomCar2005}
{Zomorodian}, A. \& {Carlsson}, G. (2005).
\newblock Computing persistent homology.
\newblock {\em Discrete and Computational Geometry}, 33, 249--274.

\end{thebibliography}

%
%
\end{document}